\documentclass[11pt]{article}

\usepackage[final]{acl}

\usepackage{times}
\usepackage{latexsym}
\usepackage[T1]{fontenc}
\usepackage[utf8]{inputenc}
\usepackage{microtype}
\usepackage{inconsolata}
\usepackage{graphicx}

\usepackage{amsmath}
\usepackage{amssymb}
\usepackage{bbm}
\usepackage{booktabs}
\usepackage{multirow}
\usepackage{wrapfig}
\usepackage{enumitem}
\usepackage{threeparttable}
\usepackage{pifont}
\usepackage{colortbl}
\usepackage{tcolorbox}
\tcbuselibrary{skins,breakable}
\usepackage{xspace}

\newcommand{\ci}[1]{{\scriptsize\,(#1)}}
\definecolor{bandblue}{HTML}{CFDDF2}
\definecolor{bandtext}{HTML}{1F3864}
\newcommand{\sh}{\cellcolor{gray!15}}
\newcommand{\blockhdr}[2]{%
  \rowcolor{bandblue}[\tabcolsep][\tabcolsep]%
  \multicolumn{9}{l}{\textcolor{bandtext}{\textbf{#1}\,---\,\textit{#2}}}}

\newcommand{\ours}{GRASP\xspace}

\newcommand{\draftonly}[1]{}

\hypersetup{
  pdftitle={GRASP: Graph Agentic Search over Propositions for Multi-hop Question Answering},
  pdfauthor={J. Stockton Jenkins, Ramya Korlakai Vinayak, Junjie Hu}
}

\title{GRASP: Graph Agentic Search over Propositions for\\Multi-hop Question Answering}

\author{J. Stockton Jenkins \qquad Ramya Korlakai Vinayak \qquad Junjie Hu \\
  University of Wisconsin--Madison \\
  \texttt{\{jsjenkins4, korlakaivina, junjie.hu\}@wisc.edu}}

\begin{document}
\maketitle

\begin{abstract}
Agentic retrieval improves multi-hop question answering by giving language models autonomy to iteratively gather evidence. Recent work augments these systems with knowledge graphs for structured traversal, but this combination introduces significant cost: expensive graph construction at index time and compounding token usage at inference time. We introduce \textbf{Gr}aph \textbf{A}gentic \textbf{S}earch over \textbf{P}ropositions (GRASP), which decomposes multi-hop questions into dependency-aware plans and scales the number of sub-agents to question complexity. Each sub-agent resolves its single-hop query over a novel three-layer hierarchical graph of entities, propositions, and passages, using the entity layer for targeted traversal and the proposition layer for high-recall passage retrieval. We evaluate on MuSiQue, 2WikiMultihopQA, and HotpotQA in two settings, open-corpus retrieval and extended-context reasoning (LongBench), under both a closed and an open model. GRASP achieves the highest accuracy on five of six benchmarks and is the most token-efficient method in every configuration. Its lead widens under the weaker open model, where baselines that rely on the LLM to navigate the graph degrade sharply. Finally, we introduce \emph{success economy}, the difficulty-weighted token cost per correct answer, and advocate for efficiency-aware evaluation as standard practice for agentic search and question-answering. Code and data are available at \url{https://github.com/stocktonjenkins/grasp}.
\end{abstract}

\section{Introduction}
\label{sec:introduction}

Large language models encode vast world knowledge, yet
retrieval-augmented generation remains essential for grounding their
outputs in verifiable, up-to-date evidence \citep{gao2024ragsurvey, lewisrag}. Single-shot retrieval is insufficient for multi-hop questions, as each hop introduces dependencies that semantic similarity alone cannot resolve \citep{trivedi2022musique,  press2023, singh2025agentragsurvey}. Agentic search addresses this by giving models autonomy to iteratively search, reason, and accumulate evidence \citep{jin2025searchr1trainingllmsreason,jiang-etal-2025-s3, li2026semanticsimilarityrethinkingretrieval}. Recent work augments these agents with knowledge graphs for structured multi-hop traversal \citep{li2024graphreader, shen2025gear}. However, this introduces a fundamental tension: \textbf{as questions become more complex, token usage grows rapidly}, driven by repeated LLM calls and expanding context across steps. This creates a bottleneck for deploying agentic systems in practice. %

\begin{figure*}[t] %
    \centering
    \includegraphics[width=\textwidth]{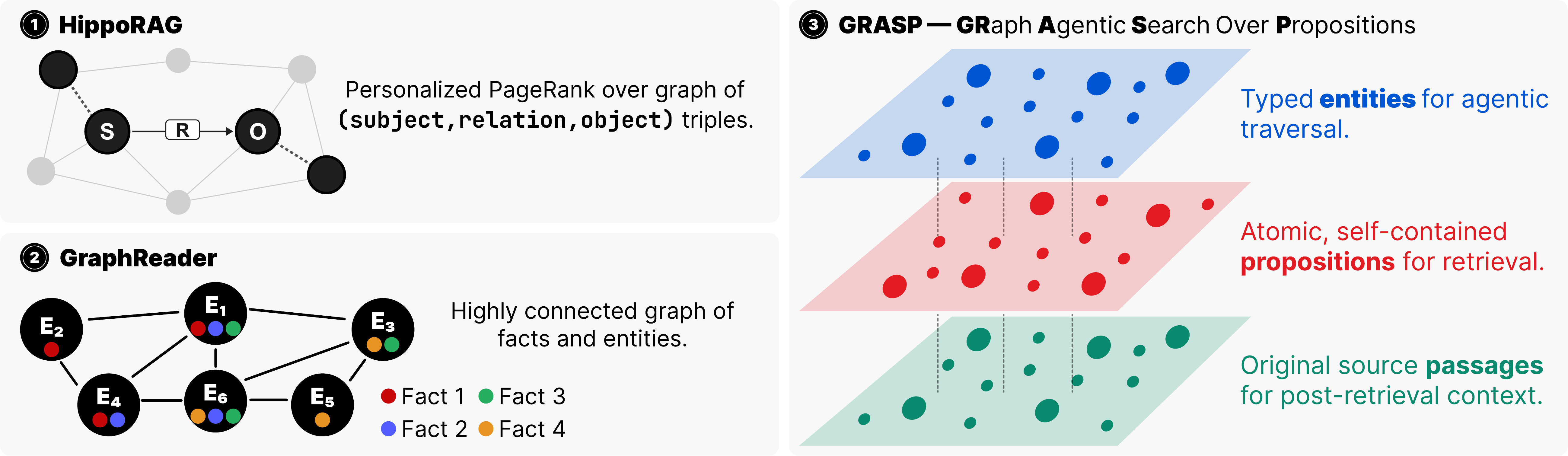}
    \caption{A comparison of graph structures for multi-hop QA. \ding{182}~\textbf{Relational knowledge graphs}: HippoRAG extracts \texttt{(subject, relation, object)} triples into a flat knowledge graph and retrieves passages via Personalized PageRank. \ding{183}~\textbf{Factual graphs}: GraphReader organizes atomic facts under key-element nodes into a highly connected graph, deployed by $N$ parallel agents for traversal. \ding{184}~\textbf{Proposition graph (ours)}: GRASP builds a hierarchical graph with three layers: typed entities for agentic traversal, propositions for retrieval, and source passages for post-retrieval context.}
    \label{fig:comparison-figure}
\end{figure*}

Existing graph-based approaches to multi-hop QA fall along three tracks. First,
\emph{retrieval-augmented} methods such as HippoRAG \citep{gutierrez2025from, gutierrez2024hipporagneurobiologicallyinspiredlongterm}
and GeAR \citep{shen2025gear} build triple-based knowledge graphs for structured
traversal. Recent work shows that triple extraction is both expensive
and unreliable, and often degrades retrieval quality below naive RAG
\citep{zhuang2025linearrag, han2026ragvsgraphragsystematic}\footnote{We build on this work and further show in \S\ref{sec:embedding-similarity-exp} that KG triples are sub-optimal retrieval units, motivating our exploration of an alternative indexing strategy.}. Second, \emph{extended-context} methods such as GraphReader
\citep{li2024graphreader} organize long texts into navigable graphs at runtime that are explored by
multiple parallel agents. Third, \emph{KG-traversal} methods \citep{sun2024thinkongraphdeepresponsiblereasoning, luo2024reasoninggraphsfaithfulinterpretable, liu2024kelp, li2025simple} prompt LLMs to reason over externally provided knowledge graphs, assuming an index rather than constructing one
from the corpus. Across all three tracks, no existing work optimizes for total
token usage for both indexing and inference in the agentic setting.

In this paper, we propose GRASP: \textbf{Gr}aph \textbf{A}gentic \textbf{S}earch
over \textbf{P}ropositions, an agentic retrieval system that improves the accuracy and token efficiency of multi-hop question answering. Motivated by our empirical observations, GRASP couples two designs. First, a novel three-layer hierarchical graph of entities, propositions, and passages, built by joint extraction that avoids relational triples. Propositions are more information-dense than triples and more precise
than full passages for a retriever, shared entities provide targeted gateways for traversal, and source passages supply full context for answer generation. Second, \emph{dependency-aware decomposition}: GRASP decomposes multi-hop questions into dependency-ordered sub-questions, each resolved independently against the graph. We find that agentic decomposition and proposition retrieval are interdependent — substituting either degrades both accuracy and token efficiency. Furthermore, resolving each sub-question in isolation further bounds context growth, so token usage stays controlled as reasoning depth increases.

In summary, we make the following contributions and findings:

\begin{enumerate}[leftmargin=20pt]
\item \textbf{A novel three-layer graph for multi-hop retrieval and QA.} We introduce a novel hierarchical graph of entities, propositions, and passages, built without relational triples. With propositions as the core retrieval unit, GRASP achieves higher passage recall than HippoRAG2's triple graph at 69\% fewer indexing tokens, extending proposition retrieval beyond the single-hop setting.

\item \textbf{Proposition retrieval and agentic decomposition are mutually reinforcing.} We design a search agent that decomposes multi-hop queries into dependency-aware plans and observe that this decomposition improves proposition-level recall in the multi-hop setting, widening GRASP's advantage over HippoRAG2. A cross-ablation of graph structure against inference architecture confirms the coupling runs both ways, as substituting either component degrades both QA accuracy and token efficiency.

\item \textbf{Efficiency-aware evaluation across open and closed models.} We evaluate on MuSiQue, 2Wiki, and HotpotQA in the retrieval and extended-context settings, using \texttt{Qwen3-32B} and \texttt{Gemini-3-flash} backbones. To our knowledge, GRASP is the first system evaluated competitively across both settings. We report token usage for every method and introduce \emph{success economy}, a metric inspired by information theory that amortizes token cost over difficulty-weighted correct answers. GRASP attains the highest accuracy on five of six benchmarks and the best success economy in every evaluated configuration.
\end{enumerate}

\section{Related Work}
\label{sec:related_work}

\textbf{Graph structures for retrieval.}
Graph-based retrieval captures inter-document relationships that flat indexing misses. HippoRAG \citep{gutierrez2024hipporagneurobiologicallyinspiredlongterm} and HippoRAG2 \citep{gutierrez2025from} extract relational triples via OpenIE and retrieve passages through personalized PageRank (Figure~\ref{fig:comparison-figure}). GeAR \citep{shen2025gear} adds LLM-guided beam search over a similar triple graph. LightRAG \citep{guo2025lightrag} proposes dual-level entity-and-relationship indexing, though it still relies on relation extraction. However, recent work \citep{zhuang2025linearrag, han2026ragvsgraphragsystematic} demonstrates that triple extraction is lossy---relations are context-dependent and difficult to compress into atomic triples without losing semantic nuance. GraphRAG \citep{edge2025localglobalgraphrag} and RAPTOR \citep{sarthi2024raptorrecursiveabstractiveprocessing} take summarization-based approaches, building hierarchical indices via community detection or recursive abstraction.

\textbf{Agentic question answering.} \cite{yao2023reactsynergizingreasoningacting, park2023generativeagentsinteractivesimulacra} established the paradigm of interleaving reflection, tool use, and memory in LLM agents. IRCoT \citep{trivedi2023ircot} interleaves chain-of-thought with retrieval; Self-Ask \citep{press2023} decomposes questions into sequential sub-questions, a strategy we adopt in our planning module. Several methods couple agents with knowledge graph traversal: Think-on-Graph \citep{sun2024thinkongraphdeepresponsiblereasoning} and Reasoning on Graphs \citep{luo2024reasoninggraphsfaithfulinterpretable} prompt LLMs to explore existing KGs, while KGP \citep{wang2023knowledgegraphpromptingmultidocument} constructs a graph from document collections. In the extended-context setting, GraphReader \citep{li2024graphreader} links key elements through shared atomic facts (Figure~\ref{fig:comparison-figure}) and deploys $N$ parallel agents, but duplicates facts across nodes and lacks inter-agent coordination, resulting in redundant traversal and wasteful token consumption. ReadAgent \citep{lee2024readagent} uses a gist memory index for selective recall without graph structure.

\textbf{Proposition-based representations.}
FActScore \citep{min2023factscorefinegrainedatomicevaluation} introduced atomic fact decomposition for evaluating factual precision. PropSegmEnt \citep{chen-etal-2023-propsegment} provides a corpus for proposition segmentation and entailment, WiCE \citep{kamoi2023wicerealworldentailmentclaims} extends claim-level entailment to Wikipedia verification, and \citet{chen2023subsentenceencodercontrastivelearning} train sub-sentence encoders for proposition-level embeddings. \citet{chen2023dense} adapted propositions to retrieval, showing that proposition-level indexing outperforms passage and sentence granularity for open-domain QA. \citet{li2024graphreader} employ atomic facts in their graph, which are similar to propositions in that they are also atomic, natural language statements.

\section{Method}
\label{sec:method}

\begin{figure*}[t]
  \centering
  \includegraphics[width=\textwidth]{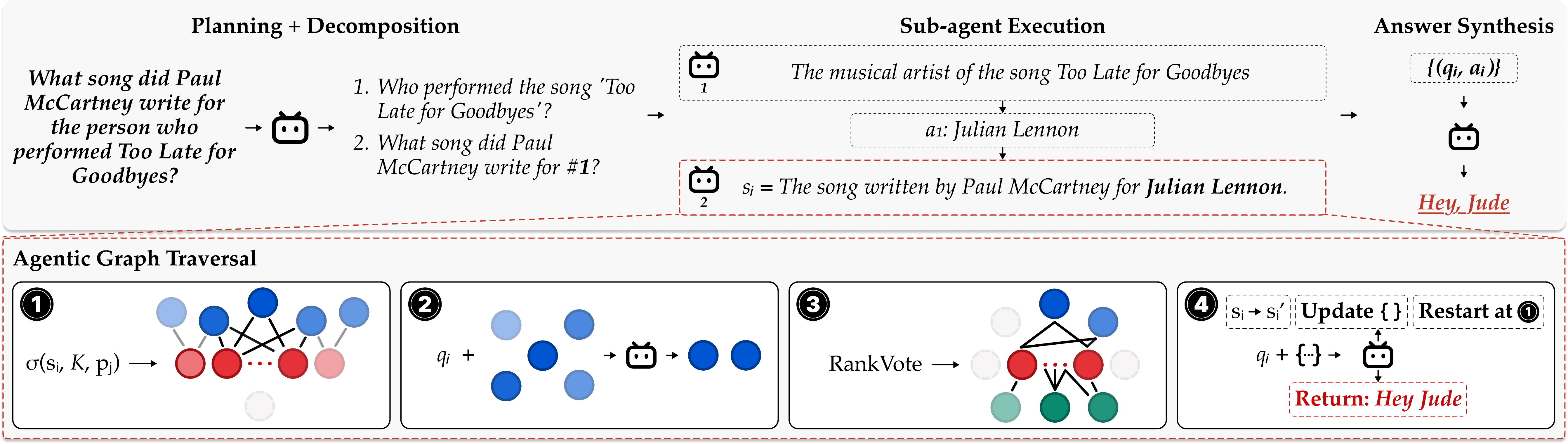}
  \caption{GRASP inference pipeline. \textbf{Top}: the planner decomposes the input into dependency-ordered sub-questions. A synthesis module produces the final answer from all sub-question--answer pairs. \textbf{Bottom}: agentic graph traversal for a single sub-agent. \ding{182}~Retrieve propositions via semantic search and aggregate to connecting entities. \ding{183}~Select relevant entities. \ding{184}~Score passage evidence by reciprocal rank vote. \ding{185}~Return answer or rewrite the search term and restart at \ding{182}. Entity, proposition, and passage nodes are colored blue, red, and green, respectively.}
  \label{fig:architecture}
\end{figure*}

As illustrated in Figure~\ref{fig:architecture}, GRASP decomposes a multi-hop question into dependency-ordered sub-questions, each resolved by a sub-agent that traverses a three-layer graph of entities, propositions, and passages with iterative state updates, before a synthesis module produces the final answer.

\subsection{Graph Construction}
\label{sec:graph_construction}

GRASP represents a corpus as a three-layer graph $\mathcal{G} = (\mathcal{V}_E, \mathcal{V}_P, \mathcal{V}_D, \mathcal{E})$, where $\mathcal{V}_E$, $\mathcal{V}_P$, and $\mathcal{V}_D$ denote entity, proposition, and passage nodes, respectively. $\mathcal{E}$ denotes the edges that connect entities to the propositions from which they were extracted  and propositions to their source passages. We visualize GRASP's hierarchical structure in Figure~\ref{fig:comparison-figure}.

Propositions and entities are extracted jointly in a single LLM call (Figure~\ref{fig:extraction}). We prompt the LLM to decompose the input into atomic, self-contained propositions and also identify named entities with typed labels, linking each entity to its source propositions via index references. These references directly form the entity--proposition edges in $\mathcal{G}$. The prompt operates over arbitrary input length, and each extracted proposition retains a link to its source passage. Entity nodes are deduplicated by canonical name and semantic type similarity, preventing graph fragmentation from minor type variation across passages. All propositions are embedded by a pre-trained language encoder for hybrid search at inference time. We show in \S\ref{sec:experiments} that proposition extraction provides a significant boost in recall over raw sentences, justifying the extraction cost.

\subsection{Agent Planning}
\label{sec:agent-planning}

Given a multi-hop question, GRASP produces a dependency-aware plan and a sequence of sub-questions. The plan is a brief outline of the reasoning chain, while the sub-questions decompose the original question into single-hop retrieval tasks, each targeting a specific fact. Furthermore, we adapt the output format to model inter-question dependencies explicitly. Sub-questions reference the answers of prior sub-questions through numeric placeholders (e.g., \textit{Where was \#1 born?}), similar to the dependency structure of MuSiQue.  The planner is implemented as a single LLM call with few-shot examples covering linear, branching, and comparison questions (Figure~\ref{fig:question-types}).

\subsection{Sub-Agent Execution}
\label{sec:subagent}

The bottom half of Figure~\ref{fig:architecture} visualizes the life cycle of each sub-agent. Each sub-question $q_i$ from the plan is resolved by an independent sub-agent $\alpha_i$, conditioned on three inputs: the overarching multi-hop question $Q$, the sub-question $q_i$, and the history of preceding question--answer pairs $\mathcal{H}=\{(q_1,a_1),\dots,(q_{i-1},a_{i-1})\}$. From these, $\alpha_i$ rewrites $q_i$ into a search term $s_i$ for dense retrieval and a keyword list $\mathcal{K}_i$ for exact entity matching.

Candidate entities are surfaced from two sources. First, we aggregate dense retrieval to the entity layer. The system retrieves the subset $\mathcal{P}_m \subset \mathcal{V}_P$ of the $m$ propositions most semantically similar to $s_i$. Each proposition is linked to the entities it mentions, so the similarity scores are aggregated and induce a ranking over entities:
\begin{equation}
  \mathrm{score}(e)
    = \frac{\sum_{p \in \mathcal{P}_m(e)}
            \cos\!\left(\mathbf{v}_{s_i}, \mathbf{v}_{p}\right)}
           {\sqrt{1 + g(e)}},
  \label{eq:entity_score}
\end{equation}
where $\mathcal{P}_m(e) \subseteq \mathcal{P}_m$ are the retrieved propositions linked to entity $e$, $g(e)$ is the degree of $e$ in $\mathcal{G}$, and $\cos(\mathbf{v}_{s_i}, \mathbf{v}_{p})$ is the cosine similarity between the embedding of search term $s_i$ and proposition $p$. The denominator dampens hub entities, which would otherwise accumulate score simply by appearing in many propositions. Second, an exact name match is strong evidence of relevance, so we also match $\mathcal{K}_i$ against entity names directly, $\mathcal{M}_i = \{e \in \mathcal{V}_E : \mathrm{name}(e) \in \mathcal{K}_i\}$. The two sources are pooled, deduplicated, and ranked together. The $k$ highest-ranked entities form the candidate set $\mathcal{C}_k \subset \mathcal{V}_E$. The sub-agent $\alpha_i$ then compares $\mathcal{C}_k$ against $q_i$ and selects a subset $\mathcal{S} \subseteq \mathcal{C}_k$ for traversal. This step gives $\alpha_i$ control over the direction of traversal and bounds the number of propositions entering the passage-level vote (see \S\ref{sec:graph-structural-ablation} for empirical justification).

As shown in step \ding{184} of Figure~\ref{fig:architecture}, sub-agent $\alpha_i$ retrieves passage-level evidence by using $s_i$ to perform a semantic search over all propositions linked to entities in $\mathcal{S}$, i.e., $\bigcup_{e \in \mathcal{S}} \mathcal{P}(e)$. Passages are scored using the \texttt{RankVote}~\citep{geirhos2025flexibleperceptionvisualmemory} of their constituent propositions. Because each proposition $p$ belongs to exactly one parent passage $D$, a passage's total score is computed by summing the reciprocal-rank votes of its connected propositions: 

\begin{equation}
    \text{score}(D) = \sum_{p \in \mathcal{P}(D)} \frac{1}{1 + R(p)},
    \label{eq:rr}
\end{equation}

where $\mathcal{P}(D)$ is the set of propositions extracted from $D$ and $R(p)$ is the proposition's rank. We score each passage by this function and return the top-$d$ passages to the sub-agent as new evidence. 

Instead of tracking a continuous message history across its entire life cycle, sub-agent $\alpha_i$ is equipped with a compact observation state $O_i$. At each LLM call, $\alpha_i$ processes its state alongside freshly retrieved evidence to generate concise, entity-level observations that are appended back to $O_i$. Furthermore, $\alpha_i$ filters previously visited entities and propositions from subsequent retrieval, ensuring $\alpha_i$ does not revisit evidence. This design bounds context growth at two levels. Within a sub-agent, $O_i$ replaces a full conversation history; each entry is a concise summary, and previously retrieved evidence is never re-read. Across sub-agents, only $Q$ and $\mathcal{H}$ propagate, discarding all tool calls, retrieved evidence, and intermediate reasoning traces.

After observing both $O_i$ and new evidence, $\alpha_i$ selects one of two actions (step~\ding{185}, Figure~\ref{fig:architecture}). If the evidence is sufficient, it emits \textbf{DONE} and returns the answer $a_i$; otherwise it emits \textbf{QUERY\_AGAIN}, rewriting the search term $s_i \rightarrow s_i'$, appending new observations to $O_i$ and restarting traversal at step~\ding{182}. After all sub-agents complete, $\mathcal{H}$ is passed to a synthesis module that produces the final answer. This module consolidates the sub-answers into a single coherent response.

\section{Experimental Setup}
\label{sec:setup}

\textbf{Datasets.} We evaluate on three widely used multi-hop reasoning datasets: MuSiQue \citep{trivedi2022musique}, 2WikiMultihopQA 
\citep{ho2020constructing}, and HotpotQA \citep{yang2018hotpotqa}. These benchmarks have been adopted by recent graph-based and agentic QA systems, including HippoRAG2, IRCoT, GeAR, GraphReader, and LinearRAG.

Each dataset originally provides per-question passage sets containing both gold supporting passages and distractors. We evaluate in two settings that repurpose these datasets differently. In the \textit{retrieval setting}, passages are pooled into a shared index and systems must retrieve relevant context at inference time. We use the HippoRAG2 splits, each containing 1{,}000 question-passage-set pairs with ground-truth passage labels and final answers. In the \textit{extended context setting}, we use the LongBench \citep{bai2024longbench} variants of the same datasets, where ground-truth passages are distributed among distractors in a single long input. Each LongBench split contains $200$ questions. Full statistics are provided in Table~\ref{tab:dataset_stats}. 

\textbf{Baselines.} We compare against methods that grant the LLM autonomy over retrieval or reasoning \emph{and} construct their own index over the corpus before inference, so that all systems are evaluated as agentic architectures under the same assumptions. \S\ref{sec:grasp-vs-others} compares GRASP to included methods as well as other related work we considered but ultimately excluded. Non-agentic, single-pass RAG baselines are reported separately in Table~\ref{tab:non_agentic}.

In the \textit{retrieval setting}, we compare against IRCoT+HippoRAG2 \citep{trivedi2023ircot,gutierrez2025from}, which interleaves chain-of-thought retrieval with a triple-based graph retriever, and GeAR \citep{shen2025gear}, which uses LLM-guided beam search over a triple graph. In the \textit{extended context setting}, we use GraphReader \citep{li2024graphreader} with $N{=}5$ parallel agents, a single-agent variant GraphReader ($N{=}1$) to isolate the effect of parallelism, and ReadAgent \citep{lee2024readagent}, which uses gist memory for selecting passages.

\textbf{Models.} All methods use the same backbone and encoder within a given run. We evaluate one open and one closed configuration: \texttt{Qwen3-32B} paired with \texttt{bge-large-en-v1.5}, and \texttt{gemini-3-flash} paired with \texttt{gemini-embedding-001}. Backbones are used for both index construction and inference, and all reported token counts include both stages. Further implementation details are found in \S\ref{sec:implementation}.

\textbf{Evaluation metrics.} We evaluate answer quality with Exact Match (EM) and token-level F1, following HippoRAG2 and GeAR. To quantify efficiency, let $\mathcal{T}_i$ denote the total token usage for question $Q_i$, aggregated across all LLM calls during indexing and inference. We report:

\textit{Total token usage.}
The sum $\sum_{i=1}^{N} \mathcal{T}_i$ across all questions, capturing end-to-end cost.

\textit{Success economy.} To capture token efficiency, we amortize token usage
over correct answers, weighted by difficulty:
\begin{equation}
\label{eq:success-economy}
C_w = \frac{\sum_{i=1}^{N} \mathcal{T}_i}
           {\sum_{i=1}^{N} w_i \, \mathbbm{1}[\text{EM}_i = 1]}
\end{equation}
where $\mathbbm{1}[\text{EM}_i = 1]$ credits $Q_i$ on an exact match and
$w_i \geq 0$ weights it by difficulty. $C_w$ penalizes token
waste, low accuracy, and accuracy concentrated on easy questions.

For $w_i$ we use \textit{surprisal} \citep{6773024}: the self-information $w_i = -\log_2 r_i$ quantifies the bits of information gained from observing a correct answer to $Q_i$, where $r_i$ is estimated by prompting the base LLM with $Q_i$ $n$ times at non-zero temperature and computing the fraction of correct answers. This assigns higher weight to questions the model rarely solves, following the principle that not all evaluation items are equally informative \citep{rodriguez-etal-2021-evaluation}. LLM sampling is not required by the metric, as any non-negative difficulty signal may be substituted, including hop count, human annotation, or uniform weights ($w_i = 1$), under which $C_w$ reduces to tokens per correct answer. \footnote{We estimate $w_i$ with LLM sampling because questions the backbone already answers from parametric memory receive near-zero weight, so systems are not credited for retrieval they did not need.} In \ref{sec:success-economy-judge} we show that recomputing the weights with a different LLM leaves the ranking of methods unchanged, so method rankings are robust to the choice of difficulty LLM judge.

Since each $r_i$ is estimated from $n$ samples, we report a $95\%$ confidence interval on $C_w$ that reflects this sampling error. Over $B = 1{,}000$ parametric bootstrap iterations, we redraw $r_i^\star = k_i^\star / n$ with $k_i^\star \sim \text{Binomial}(n, \hat{r}_i)$, where $\hat{r}_i$ is the fraction of correct answers, and recompute $C_w$ holding the EM indicators and token counts fixed.

\begin{table*}[t]
  \centering
  \setlength{\tabcolsep}{4pt}
  \renewcommand{\arraystretch}{0.95}
  \small
  \resizebox{\textwidth}{!}{%
  \begin{tabular}{l cc c l cc c l}
    \toprule
    & \multicolumn{4}{c}{\textbf{\textsc{Gemini-3-flash}}}
    & \multicolumn{4}{c}{\textbf{\textsc{Qwen3-32B}}} \\
    \cmidrule(lr){2-5} \cmidrule(lr){6-9}
    \textbf{Method}
      & EM & F1 & Tok\,(M) & $C_w\downarrow$
      & EM & F1 & Tok\,(M) & $C_w\downarrow$ \\
    \midrule
    \blockhdr{MuSiQue}{Retrieval setting} \\
    GRASP           & \textbf{0.505} & \textbf{0.651} & \textbf{15.89} & \textbf{12.6}\ci{12.5--12.7}
                          & \textbf{0.361} & \textbf{0.463} & \textbf{18.71} & \textbf{15.9}\ci{15.9--16.1} \\
    IRCoT + HippoRAG2     & 0.478 & 0.604 & 31.49 & 26.7\ci{26.4--26.8}
                          & 0.316 & 0.435 & 38.13 & 37.6\ci{37.4--37.9} \\
    GeAR                  & 0.414 & 0.530 & 16.69 & 15.5\ci{15.4--15.6}
                          & 0.308 & 0.405 & 19.17 & 19.4\ci{19.3--19.6} \\
    \blockhdr{MuSiQue}{Extended context setting} \\
    GRASP           & \textbf{0.500} & \textbf{0.625} & 9.72 & \textbf{51.9}\ci{50.4--52.9}
                          & \textbf{0.470} & \textbf{0.559} & 17.35 & \textbf{58.7}\ci{58.0--59.5} \\
    GraphReader           & 0.480 & 0.619 & 14.07 & 80.9\ci{78.5--82.7}
                          & 0.295 & 0.407 & 18.97 & 102.7\ci{100.1--105.0} \\
    GraphReader ($N{=}1$) & 0.400 & 0.538 & \textbf{9.44} & 73.7\ci{71.1--75.9}
                          & 0.155 & 0.246 & \textbf{14.60} & 207.3\ci{195.5--218.6} \\
    ReadAgent             & 0.355 & 0.466 & 14.07 & 116.7\ci{112.2--120.1}
                          & 0.270 & 0.365 & 15.00 & 91.3\ci{89.5--93.4} \\
    \blockhdr{2Wiki}{Retrieval setting} \\
    GRASP           & \textbf{0.737} & \textbf{0.825} & 10.90 & \textbf{9.5}\ci{9.3--9.5}
                          & \textbf{0.628} & \textbf{0.711} & 14.09 & \textbf{10.0}\ci{9.8--10.1} \\
    IRCoT + HippoRAG2     & 0.717 & 0.805 & 21.52 & 20.2\ci{19.8--20.4}
                          & 0.592 & 0.674 & 27.47 & 21.5\ci{20.9--21.6} \\
    GeAR                  & 0.444 & 0.478 & \textbf{9.89} & 13.7\ci{13.4--13.9}
                          & 0.382 & 0.430 & \textbf{13.87} & 13.1\ci{12.8--13.2} \\
    \blockhdr{2Wiki}{Extended context setting} \\
    GRASP           & \textbf{0.700} & \textbf{0.806} & 5.20 & \textbf{23.9}\ci{23.2--24.7}
                          & \textbf{0.615} & \textbf{0.725} & 9.86 & \textbf{37.1}\ci{35.8--38.3} \\
    GraphReader           & 0.590 & 0.764 & 8.84 & 53.3\ci{51.1--55.3}
                          & 0.310 & 0.542 & 11.05 & 72.1\ci{69.8--75.2} \\
    GraphReader ($N{=}1$) & 0.580 & 0.741 & \textbf{4.96} & 32.9\ci{31.6--34.3}
                          & 0.135 & 0.255 & 7.56 & 113.5\ci{110.0--120.8} \\
    ReadAgent             & 0.520 & 0.654 & 6.08 & 39.3\ci{37.8--40.7}
                          & 0.435 & 0.538 & \textbf{6.56} & 40.7\ci{38.7--42.5} \\
    \blockhdr{HotpotQA}{Retrieval setting} \\
    GRASP           & 0.631 & 0.772 & \textbf{12.68} & \textbf{14.5}\ci{14.3--14.6}
                          & 0.584 & 0.713 & \textbf{15.06} & \textbf{10.8}\ci{10.7--10.9} \\
    IRCoT + HippoRAG2     & \textbf{0.678} & \textbf{0.807} & 26.82 & 26.4\ci{26.1--26.5}
                          & \textbf{0.606} & \textbf{0.737} & 28.55 & 19.5\ci{19.3--19.6} \\
    GeAR                  & 0.557 & 0.677 & 13.18 & 16.4\ci{16.2--16.6}
                          & 0.508 & 0.641 & 15.76 & 13.1\ci{12.9--13.2} \\
    \blockhdr{HotpotQA}{Extended context setting} \\
    GRASP           & \textbf{0.535} & \textbf{0.681} & 8.00 & \textbf{51.6}\ci{50.2--53.3}
                          & \textbf{0.465} & \textbf{0.599} & 14.95 & \textbf{71.8}\ci{69.9--73.6} \\
    GraphReader           & 0.515 & 0.679 & 12.23 & 76.4\ci{74.0--78.0}
                          & 0.320 & 0.510 & 16.40 & 90.9\ci{89.0--93.4} \\
    GraphReader ($N{=}1$) & 0.485 & 0.640 & \textbf{7.96} & 56.4\ci{54.3--57.9}
                          & 0.250 & 0.405 & \textbf{12.33} & 96.2\ci{92.6--98.4} \\
    ReadAgent             & 0.445 & 0.612 & 11.53 & 99.8\ci{95.9--102.4}
                          & 0.430 & 0.569 & 12.40 & 72.0\ci{69.1--74.1}\\
    \bottomrule
  \end{tabular}%
  }
  \caption{Results across \texttt{gemini-3-flash}, \texttt{Qwen3-32B} in the retrieval and extended context settings on MuSiQue, 2Wiki, and HotpotQA. Best results per block are \textbf{bolded}. Tokens in millions (M). $C_w$: success economy in thousands of tokens per difficulty-weighted correct answer, with bootstrap 95\% confidence intervals in parentheses ($\downarrow$: lower is better).}
  \label{tab:main_results}
\end{table*}

\section{Experiments}
\label{sec:experiments}

\begin{table*}[t]
\centering

\begin{minipage}[t]{0.50\textwidth}
  \centering
  \footnotesize
  \setlength{\tabcolsep}{3pt}
  \renewcommand{\arraystretch}{0.95}
  \begin{tabular}{@{}llllcc@{}}
    \toprule
    \textbf{Mode} & \textbf{Method} & \textbf{Unit} & \textbf{Weight} & \textbf{Recall} & \textbf{Idx.} \\
    \midrule
    \multirow{5}{*}{\shortstack[l]{Sim.\\Agent}}
    & HippoRAG2 & -- & -- & 0.818 & 17.6M \\
    \cmidrule(l){2-6}
    & GRASP$^*$ & Prop. & RankVote & \textbf{0.904} & \multirow{2}{*}{5.4M} \\
    & GRASP     & Prop. & Uniform  & 0.848 & \\
    & GRASP     & Sent. & RankVote & 0.853 & \multirow{2}{*}{3.3M} \\
    & GRASP     & Sent. & Uniform  & 0.648 & \\
    \midrule
    \multirow{5}{*}{\shortstack[l]{Single\\Pass}}
    & HippoRAG2 & -- & -- & 0.702 & 17.6M \\
    \cmidrule(l){2-6}
    & GRASP     & Prop. & RankVote & \textbf{0.729} & \multirow{2}{*}{5.4M} \\
    & GRASP     & Prop. & Uniform  & 0.610 & \\
    & GRASP     & Sent. & RankVote & 0.656 & \multirow{2}{*}{3.3M} \\
    & GRASP     & Sent. & Uniform  & 0.429 & \\
    \bottomrule
  \end{tabular}
  \caption{Passage-level recall on MuSiQue, split from
  \cite{gutierrez2025from}. Sim.\ Agent retrieves
  $k_i = 5 \times n^{(i)}_{\text{hops}}$ passages per question; Single-Pass
  uses $k = 5$. Prop.: proposition; Sent.: sentence; Idx.: indexing tokens.
  $^*$Configuration used in Table~\ref{tab:main_results}.}
  \label{tab:musique_retrieval}
\end{minipage}\hfill
\begin{minipage}[t]{0.47\textwidth}
  \centering
  \footnotesize
  \setlength{\tabcolsep}{3pt}
  \renewcommand{\arraystretch}{0.95}
  \begin{tabular}{@{}cll cc cc@{}}
    \toprule
    & & & \multicolumn{2}{c}{\textbf{Accuracy}}
        & \multicolumn{2}{c}{\textbf{Efficiency}} \\
    \cmidrule(lr){4-5} \cmidrule(lr){6-7}
    & \textbf{Graph} & \textbf{Agent}
      & EM & F1 & Tok & $C_w$ $\downarrow$ \\
    \midrule
    \multirow{4}{*}{\rotatebox[origin=c]{90}{\scriptsize Retrieval}}
      & \sh Prop. & \sh GRASP & \sh\textbf{0.505} & \sh\textbf{0.651} & \sh\textbf{15.89} & \sh\textbf{12.6} \\
      & Triple & GRASP & 0.496 & 0.638 & 29.12 & 23.8 \\
      & Prop.  & IRCoT & 0.401 & 0.547 & 19.25 & 18.2 \\
      & \sh Triple & \sh IRCoT & \sh 0.478 & \sh 0.604 & \sh 31.49 & \sh 26.7 \\
    \midrule
    \multirow{4}{*}{\rotatebox[origin=c]{90}{\scriptsize Long-ctx}}
      & \sh Prop. & \sh GRASP & \sh\textbf{0.500} & \sh\textbf{0.625} & \sh\textbf{9.72} & \sh\textbf{51.9} \\
      & Prop. & GraphReader & 0.430 & 0.562 & 10.64 & 72.0 \\
      & \sh Key-elem. & \sh GraphReader & \sh 0.480 & \sh 0.619 & \sh 14.07 & \sh 80.9 \\
      & Key-elem. & GRASP & 0.495 & 0.606 & 18.16 & 93.7 \\
    \bottomrule
  \end{tabular}
  \caption{Cross-ablation of graph structure and inference agent on MuSiQue.
  Prop.: proposition graph; Triple: HippoRAG triple graph; Key-elem.:
  GraphReader key-element graph. Tok: total tokens (M).
  \colorbox{gray!15}{Shaded} rows are from Table~\ref{tab:main_results};
  unshaded rows swap the graph or traversal method.}
  \label{tab:cross_ablation}
\end{minipage}

\end{table*}

\textbf{Question answering evaluation.} Table~\ref{tab:main_results} reports
results for all configurations.

\textit{Retrieval setting.} GRASP scores the highest EM and F1 on MuSiQue and 2Wiki under both backbones, outperforming IRCoT+HippoRAG2 on both metrics while using roughly \textbf{50\% fewer total tokens}. On HotpotQA, IRCoT+HippoRAG2 leads on accuracy under both backbones, but at roughly $2\times$ the token cost; in \S\ref{sec:hotpot-integrity} we explain why HotpotQA in the retrieval setting is the least informative dataset in our experiments. GeAR has similar token usage to GRASP but trails substantially on accuracy, falling 29 EM points behind on 2Wiki under Gemini. Despite its low cost, it yields worse success economy on every dataset. \textbf{Critically, GRASP attains the best success economy on all three datasets under both backbones}, indicating that it is more token efficient for correctly answering hard questions.

\textit{Extended context setting.} GRASP leads on EM and F1 across all three datasets under both backbones. Under Gemini it does so while using 30--40\% fewer tokens than GraphReader. Under Qwen the token gap narrows to under 15\%, yet GRASP's success-economy advantage widens because GraphReader's accuracy collapses. \textbf{GRASP has the lowest success economy in all twelve dataset--setting--backbone blocks and is statistically separable from the runner-up in eleven}. The exception is HotpotQA under Qwen, where ReadAgent matches GRASP's $C_w$ by answering cheaply though less accurately. Interestingly, we observe that the methods that ask the LLM to \textit{navigate} the graph degrade most as the backbone weakens. GraphReader navigates through repeated LLM calls without embedding-based retrieval, so its accuracy depends on model capability. This is further corroborated with GraphReader $N{=}1$. Removing GraphReader's parallelism collapses its accuracy further under Qwen, to 0.155 EM on MuSiQue and 0.135 on 2Wiki. GRASP anchors retrieval in embedding search over propositions and uses the LLM only to \textit{select} among candidates, reducing the inference burden from \textit{navigation} to \textit{selection}.

\textbf{Retrieval evaluation.} We benchmark GRASP's retrieval mechanism against HippoRAG2 under two configurations on the MuSiQue dataset, evaluating passage-level recall for both. We use \texttt{gemini-embedding-001} as the encoder in our retrieval experiments.

\textit{Simulated agentic retrieval.} To simulate the sub-agent retrieval process during QA in GRASP, we utilize MuSiQue's annotations to extract sub-questions. For each sub-question, GRASP retrieves the top $k=5$ passages. This setting measures the upper-bound retrieval performance of GRASP's decomposed approach. Conversely, HippoRAG2 performs a \textit{single retrieval pass} using the entire multi-hop question. To establish the fairest comparison possible and ensure both methods retrieve the exact same total number of passages, we dynamically extend HippoRAG2's retrieval limit to $k_i = 5 \times n^{(i)}_{\textbf{hops}}$ for a given question $Q_i$.

\textit{Standard single-pass retrieval.} To provide a direct and strict comparison to HippoRAG2's default evaluation setting, we restrict the total retrieved passages to $k=5$ to both methods. GRASP abandons sub-question decomposition and instead uses the original, multi-hop question $Q_i$ to query its graph in a single pass.

\textit{Retrieval results.} The results in Table~\ref{tab:musique_retrieval} demonstrate the effectiveness GRASP in retrieving ground truth passages. Under the simulated agentic setting, the full GRASP configuration achieves a recall of \textbf{0.904}, outperforming HippoRAG2's 0.818. While GRASP's retrieval mechanism was not designed for multi-hop retrieval in a single-pass, it surprisingly maintains an advantage even in this setting, achieving 0.729 compared to HippoRAG2's 0.702. 

\textit{Propositions as retrieval units.} Using extracted propositions rather than raw sentences improves recall in every configuration tested, demonstrating they are a superior retrieval unit. While indexing propositions naturally incurs a higher token cost than sentences , this trade-off is favorable; GRASP still requires 69\% fewer indexing tokens than HippoRAG2.

\textit{RankVoting.} To aggregate retrieval unit hits to passages, we use RankVoting as described in \cite{geirhos2025flexibleperceptionvisualmemory}, which outperforms uniform weighting. RankVoting allots each retrieval unit a weighted-vote by its reciprocal rank in the set. By weighting passages according to the ranks of their matched propositions, GRASP effectively prioritizes highly relevant context, yielding a nearly 6-point recall improvement over uniform weighting in the simulated agentic setting and 11-point increase in the single-pass setting.

Taken together, these results identify three design choices whose gains stack for effective multi-hop retrieval. Switching the retrieval unit from sentences to propositions raises recall from 0.429 to 0.610, reciprocal-rank voting lifts it to 0.729, and agentic decomposition carries it to 0.904. GRASP's margin over HippoRAG2 widens from 2.7 to 8.6 points once decomposition is applied, confirming that decomposition and proposition retrieval reinforce each other.

\textbf{Graph structure vs.\ inference architecture.} To disentangle the
contributions of GRASP's proposition graph from its inference architecture, we
cross each graph structure with each traversal method on MuSiQue
(Table~\ref{tab:cross_ablation}) using \texttt{gemini-3-flash}.

\textit{Retrieval setting.} Two swaps isolate the roles of the graph and the
agent. Running GRASP's agent over HippoRAG2's triple graph yields accuracy
comparable to full GRASP  but nearly doubles success economy , reflecting the cost of triple extraction. This variant still
improves over IRCoT+HippoRAG2 on both accuracy and success economy, indicating
that decomposition-based planning outperforms interleaving retrieval with
chain-of-thought regardless of the underlying graph. Swapping in the opposite
direction has the reverse effect. Running IRCoT over a proposition graph
drops accuracy to 0.401 EM, below IRCoT on its native triple graph.
IRCoT chain-of-thought-derived queries are misaligned with the proposition layer and can compound retrieval errors across CoT iterations \citep{jiang2023flare}.

\textit{Extended context setting.} Neither swap improves on GRASP. Running
GraphReader over GRASP's proposition graph lowers accuracy
relative to GraphReader on its native graph. GRASP extracts a narrow
set of named entities, whereas GraphReader's parallel explorers are co-designed
for a broader key-element vocabulary spanning nouns, verbs, and adjectives. Over
the sparser entity nodes, its explorers converge on the same gateways, duplicating each other's traversal, lowering overall coverage. GraphReader also cannot exploit the proposition layer's dense embeddings, since it traverses by key-element expansion rather than retrieval. Conversely, running GRASP's agent over GraphReader's graph produces the least efficient configuration. Without embedded nodes the agent cannot perform hybrid search and must rely entirely on LLM-driven traversal, inflating token usage with no accuracy gain over the proposition graph. Together these results show that GRASP's favorable accuracy-to-cost tradeoff comes from pairing a decomposing agent with embedding-based retrieval at the proposition layer.

\textbf{Planner Analysis.} Table~\ref{tab:planner_hop_accuracy} evaluates the planner's decomposition quality on MuSiQue by comparing the number of planned steps to the ground-truth hop count. Overall, the planner produces the correct number of steps for 82.2\% of questions, though accuracy degrades from 89.4\% at 2-hop to 66.9\% at 4-hop. When the planner matches the true hop count, EM improves by nearly 9 points, confirming that planning and decomposition quality directly impacts downstream QA accuracy. In \S~\ref{sec:grasp-err-analysis}, we perform a manual error analysis of the traces of 50 questions observe that 26\% of errors were attributable to planning.

This result also helps explain a pattern in Table~\ref{tab:cross_ablation}: despite the proposition graph achieving substantially higher retrieval recall than HippoRAG2's triple graph, the end-to-end EM gap is only 1 point. Planning errors likely impose a ceiling that limits how much better retrieval can translate into better answers. GRASP currently generates plans via few-shot examples in context. Improving planner accuracy via supervised fine-tuning or reinforcement learning with verifiable rewards is a promising direction for future work.

\begin{table}[t]
  \centering
  \small
  \setlength{\tabcolsep}{4pt}
  \begin{tabular}{lrrrr}
    \toprule
    \textbf{Hops} & $N$ & \shortstack{Plan\\Acc.}
      & \shortstack{EM\\(match)} & \shortstack{EM\\(no match)} \\
    \midrule
    2 & 518 & 89.4\% & 57.2\% & 47.3\% \\
    3 & 316 & 78.5\% & 50.0\% & 48.5\% \\
    4 & 166 & 66.9\% & 35.1\% & 32.7\% \\
    \midrule
    Overall & 1000 & 82.2\% & 52.1\% & 43.3\% \\
    \bottomrule
  \end{tabular}
  \caption{Planner hop accuracy on MuSiQue. Plan Acc.\ is the fraction of questions where the number of planned steps equals the true hop count. EM (match) and EM (no match) are exact-match scores split by whether the planner generated the correct number of steps.}
  \label{tab:planner_hop_accuracy}
\end{table}

\section{Conclusion}
\label{sec:conclusion}

We presented GRASP, a graph-enhanced agentic retrieval system that treats token efficiency as a first-class design objective alongside accuracy. By constructing a three-layer hierarchical graph from jointly extracted propositions and entities, GRASP avoids the cost and noise of relational triple extraction, reaching higher passage recall than a triple graph at 69\% fewer indexing tokens. Its planning-based inference decomposes multi-hop questions into single-hop sub-tasks handled by isolated sub-agents, bounding context growth as reasoning depth increase. We found that coupling agentic decomposition with propositional retrieval improves recall, and substituting either degrades both accuracy and efficiency. GRASP attains the highest accuracy on five of six benchmarks and the best success economy in every configuration, and its advantage grows under the weaker open model, where baselines that rely on the LLM to navigate the graph degrade sharply. We hope success economy encourages efficiency to be reported alongside accuracy as standard practice for agentic QA, where inference cost scales directly with reasoning complexity.

\section*{Limitations}
\label{sec:limitations}

\textbf{Evaluation scope.} We evaluate on three English, Wikipedia-derived multi-hop QA benchmarks. We do not evaluate on other languages or on domain-specific corpora (though we do demonstrate the effectiveness of proposition retrieval in other corpora in \S\ref{sec:embedding-similarity-exp}). Our results indicate that GRASP's relative advantage grows as the backbone weakens, but we have not tested the small-model regime.

\textbf{Errors early in the dependency chain propagate and cannot be revised.} GRASP resolves sub-questions in dependency order, conditioning each sub-agent on its predecessors' answers. A sub-agent can retry its own search via \textbf{QUERY\_AGAIN}, but once it commits an answer there is no mechanism to revisit it. A wrong intermediate answer is therefore inherited by every downstream sub-agent, which spends its full token budget on a chain that can no longer be correct. Recovering from upstream errors would require verifying a sub-answer before committing to it, or a backtracking policy over the plan.

\textbf{Success economy depends on the difficulty judge.} In our experiments, we estimate the difficulty weights $w_i$ by sampling an LLM, so absolute $C_w$ values shift with the choice of judge and should not be compared across judges. Method rankings are robust to this choice. Recomputing the weights with a second LLM leaves the ranking unchanged (§\ref{sec:success-economy-judge}), though we test this only on MuSiQue. The metric also uses total token count as a proxy for cost, which ignores latency and hardware utilization.

\section*{Acknowledgment}

This research was supported in part by the National Science Foundation under Award \#IIS-2449768 and the Technical AI Safety Research Program at Coefficient Giving. The views and conclusions expressed in this work are those of the authors and should not be interpreted as representing the official policies or endorsements of the U.S. Government, the National Science Foundation, or Coefficient Giving.

\bibliography{grasp}

\appendix
\section{Appendix}

\subsection*{Summary}

This appendix supplements the main paper by first comparing GRASP to prior work in more detail. Furthermore, we compare GRASP to non-agentic baselines, discuss HotpotQA data integrity issues, share implementation details and structural ablations validating each layer of GRASP's index. Finally, all prompts used in GRASP's agent framework are provided in Section~\ref{sec:appendix-prompts}.

\begin{figure*}[t]
  \centering
  \includegraphics[width=\textwidth]{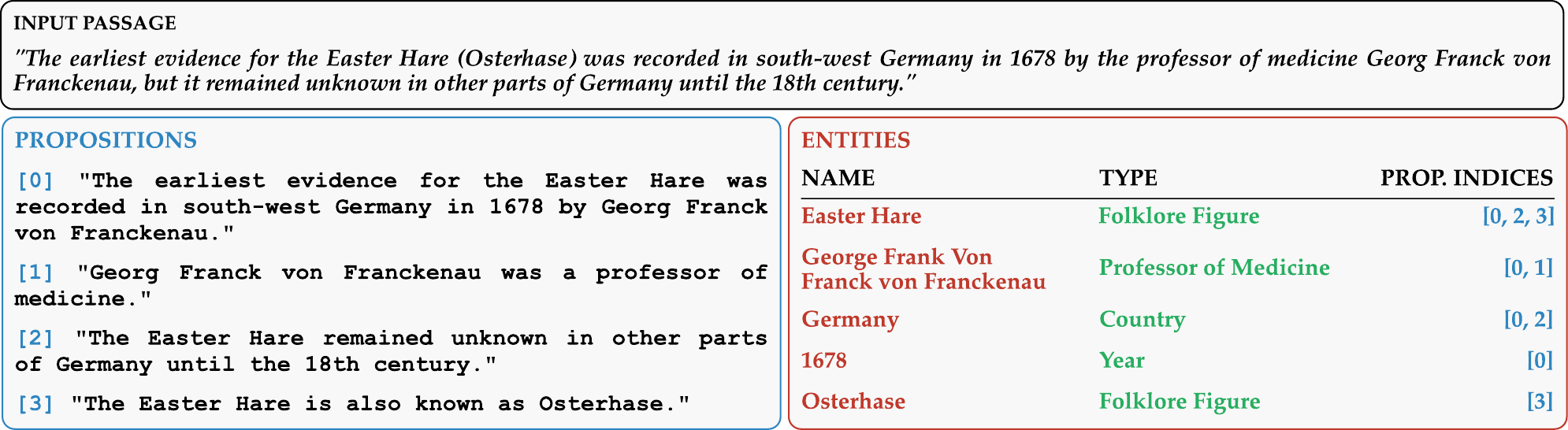}
  \caption{Proposition and entity extraction from a single passage. The passage is decomposed into propositions, with entities extracted simultaneously and linked to propositions via index references. These links form the entity--proposition edges of the graph (\S\ref{sec:graph_construction}).}
  \label{fig:extraction}
\end{figure*}

\subsection{Comparing GRASP to Prior Work}

\label{sec:grasp-vs-others}

Table~\ref{tab:baseline_design_space} compares GRASP to prior work along three axes: whether the method constructs its own \emph{LLM-built index} over the corpus, whether it employs a \emph{planning} module, and whether it performs question \emph{decomposition}. Most methods issue a single retrieval query, rely on iterative chain-of-thought, perform graph expansion without structured sub-question coordination, or even rely on a pre-built graph (e.g. Wikidata or Freeform). GraphReader is the only other method with a planning module, but without decomposition its  N parallel agents traverse independently with no shared state, producing redundant reads. KAG decomposes via logical-form translation, a fixed transformation applied upfront that does not condition on retrieved evidence; GRASP instead produces dependency-ordered natural-language sub-questions informed by prior intermediate answers. Finally, methods with no index entry (Think-on-Graph, Reasoning on Graphs, KELP, Simple is Effective) do not build their own index---they traverse externally provided KGs such as Freebase or Wikidata, a fundamentally different assumption from GRASP, which builds its index from the document corpus directly. GRASP is the only method combining all three: an LLM-built hierarchical index, a planning module, and adaptive question decomposition.

\begin{table*}[h]
\centering
\small
\setlength{\tabcolsep}{5pt}
\renewcommand{\arraystretch}{1.0}
{%
\begin{tabular}{l l c l}
\toprule
\textbf{Method} & \textbf{LLM-built Index} & \textbf{Planning} & \textbf{Question Decomposition} \\
\midrule
\textbf{GRASP (ours)} & Hier.\ proposition graph & $\checkmark$ & Sub-questions w/ dependencies \\
\midrule
IRCoT + HippoRAG2     & Triple KG              & $\times$ & --- \\
GeAR                  & Triple KG              & $\times$ & --- \\
GraphReader           & Atomic-fact graph      & $\checkmark$ & --- \\
ReadAgent             & Page-level gists       & $\times$ & --- \\
KAG                   & Triple graph           & $\times$ & Logical form (compiled) \\
LinearRAG             & Tripartite graph       & $\times$ & --- \\
LightRAG              & Knowledge graph        & $\times$ & --- \\
Think-on-Graph        & ---                    & $\times$ & --- \\
Reasoning on Graphs   & ---                    & $\times$ & --- \\
KELP                  & ---                    & $\times$ & --- \\
Simple is Effective   & ---                    & $\times$ & --- \\
\bottomrule
\end{tabular}%
}
\caption{GRASP compared to prior work along three axes: index type (constructed via an LLM), a planning module, and question decomposition. GRASP is the only method combining all three.}
\label{tab:baseline_design_space}
\end{table*}

\subsection{Implementation Details}
\label{sec:implementation}

Our primary configuration uses \texttt{gemini-3-flash-preview} with
\texttt{thinking\_level="minimal"} via the Google AI API\footnote{\url{https://ai.google.dev/gemini-api/docs}} for all LLM calls, and
generates all embeddings using \texttt{gemini-embedding-001}. To assess
robustness to the underlying backbone, we additionally evaluate an open-weight
configuration in which every LLM call, both graph construction and inference,
uses \texttt{Qwen3-32B} with thinking disabled, paired with
\texttt{BAAI/bge-large-en-v1.5} embeddings.\footnote{\texttt{Qwen3-32B}:
\url{https://huggingface.co/Qwen/Qwen3-32B}. Beyond the temperatures noted below,
we use the model's default sampling values for \texttt{top\_p}, \texttt{top\_k},
and \texttt{min\_p}.} We serve \texttt{Qwen3-32B} locally
with vLLM using tensor parallelism across two NVIDIA L40 (48\,GB) GPUs, with a
maximum context length of $32{,}768$ tokens; all other hyperparameters are held
identical across the two backbones. We use Neo4j for graph storage as an
implementation convenience, not an architectural requirement. For model
hyperparameters, we set the temperature to $0.2$ for LLM generators to allow for
slight flexibility in reasoning, and $0.1$ for LLM extractors.

During graph construction, we chunk passages into paragraph-level segments and
jointly extract propositions and entities from each chunk. On Gemini we batch ten
chunks per extraction call in both settings, and we do the same on Qwen in the
retrieval setting. In the extended-context setting, however, the longer inputs
triggered KV-cache throttling on our hardware, so we process one chunk at a time
on Qwen. We also provide few-shot examples to Qwen to improve the output's format for EM/F1 evaluation, following HippoRAG\footnote{\url{https://github.com/OSU-NLP-Group/HippoRAG/blob/legacy/src/ircot_hipporag.py}}. 

The extraction prompt is given in \S\ref{sec:joint-extraction-prompt}. When resolving entities that share a canonical name, we merge them if the cosine similarity between their respective entity types meets or exceeds a threshold of $\tau = 0.7$.

In the retrieval stage, we then score the
top $m = 50$ propositions as described in \ref{eq:entity_score} and aggregate
these scores to the entity layer, returning the top $k = 5$ candidate entities
for the sub-agent to evaluate and select from for targeted traversal. The
sub-agent receives the top $d = 5$ passages ranked by \texttt{RankVote} as new evidence. For the probability estimates $r_i$ in Eq.~\ref{eq:success-economy}, we run the evaluation backbone $n = 10$ times on each question with temperature $t = 1$ and take the fraction of correct answers; difficulty is thus estimated separately for each backbone. Finally, each sub-agent performs a maximum of $2$ iterations through steps 1--4 depicted in Figure~\ref{fig:architecture}.

\subsection{Experiments}
\label{sec:appendix-exp}
\subsubsection{Non-agentic Baselines}
\label{sec:non-agentic-baselines}

To quantify the value of agentic retrieval, we measure accuracy for four non-iterative baselines: \textbf{No Context} (zero-shot Gemini), \textbf{RAG} (single-pass retrieval with Gemini), \textbf{Self-Ask + RAG} (decomposition-based prompting from \citet{press2023} with single-pass dense retrieval per sub-question), question decomposition paired with RAG, and \textbf{Full Context} (Gemini given the complete LongBench passage set).

As shown in Table~\ref{tab:non_agentic}, GRASP consistently outperforms these baselines across all metrics. In the retrieval setting, GRASP surpasses single-pass RAG by 23.9, 37.6, and 5.2 EM points on MuSiQue, 2Wiki, and HotpotQA, respectively. Self-Ask + RAG provides a useful intermediate comparison: it improves over flat RAG on 2Wiki (+17 EM), where sub-questions are cleanly separable compositional lookups, but under performs flat RAG on MuSiQue (-3 EM). Without structured retrieval to ground each sub-question's evidence, errors in decomposition cascade through the chain. GRASP also vastly outperforms question decomposition over RAG, demonstrating that simple decomposition alone cannot replicate GRASP's performance. 

Furthermore, in the extended context setting, GRASP substantially outperforms the Full Context baseline, where Gemini is given all passages simultaneously, by 15 to 34 EM points. This demonstrates that structured, agentic graph exploration extracts and synthesizes relevant evidence far more effectively than simply expanding an LLM's context window for multi-hop reasoning problems.

\begin{table*}[t]
  \centering
  \small
  \renewcommand{\arraystretch}{1.15}
  \begin{tabular}{ll cc cc cc}
    \toprule
    & & \multicolumn{2}{c}{MuSiQue}
      & \multicolumn{2}{c}{2Wiki}
      & \multicolumn{2}{c}{HotpotQA} \\
    \cmidrule(lr){3-4} \cmidrule(lr){5-6} \cmidrule(lr){7-8}
    \textbf{Setting} & \textbf{Method} & EM & F1 & EM & F1 & EM & F1 \\
    \midrule
    \multirow{4}{*}{Retrieval}
      & No Context     & 0.178 & 0.291 & 0.396 & 0.490 & 0.395 & 0.532 \\
      & RAG            & 0.266 & 0.466 & 0.361 & 0.420 & 0.572 & 0.702 \\
      & Self-Ask + RAG & 0.332 & 0.438 & 0.653 & 0.732 & 0.575 & 0.722 \\
      & Q Dec + RAG & 0.305 & 0.395 & 0.430 & 0.492 & 0.562 & 0.699 \\
      & GRASP (Ours)   & \textbf{0.505} & \textbf{0.651} & \textbf{0.737} & \textbf{0.825} & \textbf{0.631} & \textbf{0.772} \\
    \midrule
    \multirow{2}{*}{Ext. Context}
      & Full Context   & 0.360 & 0.449 & 0.370 & 0.545 & 0.385 & 0.566 \\
      & GRASP (Ours)   & \textbf{0.500} & \textbf{0.625} & \textbf{0.700} & \textbf{0.806} & \textbf{0.535} & \textbf{0.681} \\
    \bottomrule
  \end{tabular}
  \caption{Comparison with non-agentic baselines. No Context: zero-shot
  Gemini. RAG: single-pass retrieval. Q Dec + RAG: decomposing the multi-hop query into a set of sub-questions (similar to GRASP). Retrieves top-k per sub-question and deduplicates before generation. Self-Ask + RAG: decomposition-based
  prompting (Press et al., 2023) with single-pass dense retrieval per
  sub-question. Full Context: all LongBench passages provided.}
  \label{tab:non_agentic}
\end{table*}

\subsubsection{GRASP Error Analysis}
\label{sec:grasp-err-analysis}

To diagnose where GRASP could be improved, we manually inspect the trace of 50 incorrect predictions from MuSiQue (retrieval setting) to identify where residual errors originate. We also categorize each along two axes: by primary cause (\textit{retrieval}, \textit{reasoning}, or \textit{dataset}) and by the pipeline stage at which the failure was introduced. Results are reported in Table~\ref{tab:error-analysis-by-step}.

Our analysis shows that 6 cases (12\%) are dataset-level failures in which the question targets one real-world entity but the gold answer requires a different entity of the same name (e.g., Atlanta, GA vs.\ Atlanta, MI; Francis Bacon the painter vs.\ the philosopher). The agent correctly recognizes the two entities as distinct, but the evaluation penalizes it for refusing to bridge them. Of the remaining 44 agent errors, 36 are reasoning-side and only 8 are retrieval-side --- a ratio that empirically validates our design choice in \S\ref{sec:graph_construction}: the proposition graph achieves high passage recall (Table~\ref{tab:musique_retrieval}) and is rarely the bottleneck.

The 36 reasoning-side errors split between planner errors
(13, almost entirely on 3- and 4-hop questions, consistent with the
planner-accuracy decline in Table~\ref{tab:planner_hop_accuracy}) and
sub-agent judgment errors (23) in which the retrieved evidence
and resolved context are sufficient to answer correctly but the
sub-agent reaches a wrong conclusion. We observe four recurring causes:
\textit{salience bias} (picking the most prominent among several valid
candidates in evidence), \textit{scope or granularity mismatch}
(answering at the wrong aggregation level), \textit{parametric leakage}
(overriding retrieved evidence with world knowledge), and
\textit{semantic misreading} (the sub-agent misinterprets the
sub-question and extracts an adjacent-but-wrong fact). Both the planner and sub-agent in GRASP are currently realized through prompting. Learning these control policies, as in Self-RAG and Search-R1 (\S\ref{sec:related_work}), is the natural lever for reducing the reasoning errors mentioned.

\begin{table*}[t]
\centering
\small
\begin{tabular}{lrrrrrrr}
\toprule
\textbf{Category} & \textbf{Plan} & \textbf{Step 0} & \textbf{Step 1} & \textbf{Step 2} & \textbf{Step 3} & \textbf{N/A} & \textbf{Total} \\
\midrule
Retrieval Wrong Node          & --   & 1 & -- & -- & 1 & --  & 2  \\
Retrieval Missing Fact        & --   & 1 & 3  & 1  & 1 & --  & 6  \\
Reasoning Judgment Error      & --   & 4 & 10  & 8  & 1 & --  & 23 \\
Reasoning Wrong Plan          & 13   & --& -- & -- & --& --  & 13 \\
Dataset Error                 & --   & --& -- & -- & --& 6   & 6  \\
\midrule
\textbf{Total}                & 13   & 6 & 13 & 9  & 3 & 6   & 50 \\
\bottomrule
\end{tabular}
\caption{Manual error analysis on 50 incorrect predictions, stratified by the pipeline stage at which the failure was introduced. ``Plan'' denotes errors originating in the planner's decomposition; ``Step $k$'' denotes errors made by the $k$-th sub-agent during execution; ``N/A'' marks cases where the underlying dataset --- not the agent --- is at fault (e.g.\ same-name entity collisions across cities or people). Planner errors and sub-agent mis-extraction at the middle hops (Steps~1--2) account for the bulk of failures.}
\label{tab:error-analysis-by-step}
\end{table*}

\subsubsection{Retrieval Unit Embedding Discriminability}
\label{sec:embedding-similarity-exp}

To understand why proposition-based retrieval outperforms KG triple–based methods, we investigate whether the choice of retrieval unit affects how well a query embedding can discriminate relevant units from irrelevant ones — independent of the retrieval pipeline itself. While prior work such as LinearRAG \citep{zhuang2025linearrag} has attributed KG failures to extraction errors (e.g., negation inversions), we hypothesize a more fundamental issue: even correctly extracted triples are inferior embedding targets, as compressing information into subject–predicate-object form discards contextual nuance that embedding models rely on to match natural-language queries.

To test this, we sample 1,000 questions from three single-hop retrieval benchmarks — SQuAD \citep{rajpurkar2016squad100000questionsmachine}, Natural Questions \citep{kwiatkowski-etal-2019-natural}, and PubMedQA \citep{jin2019pubmedqadatasetbiomedicalresearch} — and extract retrieval units from each question's gold passages using proposition-based extraction (GRASP), triple-based extraction (HippoRAG and GeAR), and naive sentence splitting. Sentence splitting serves as a format-controlled baseline: like propositions it produces natural-language units, but without atomic decomposition. All units from all 1,000 questions are pooled into a shared index per method; for each question, every unit in the pool is ranked by cosine similarity to the query, with gold units treated as relevant. We report NDCG@5, measuring whether units from the correct passage are surfaced near the top of the full ranking.

\begin{figure*}[t]
    \centering
    \includegraphics[width=\textwidth]{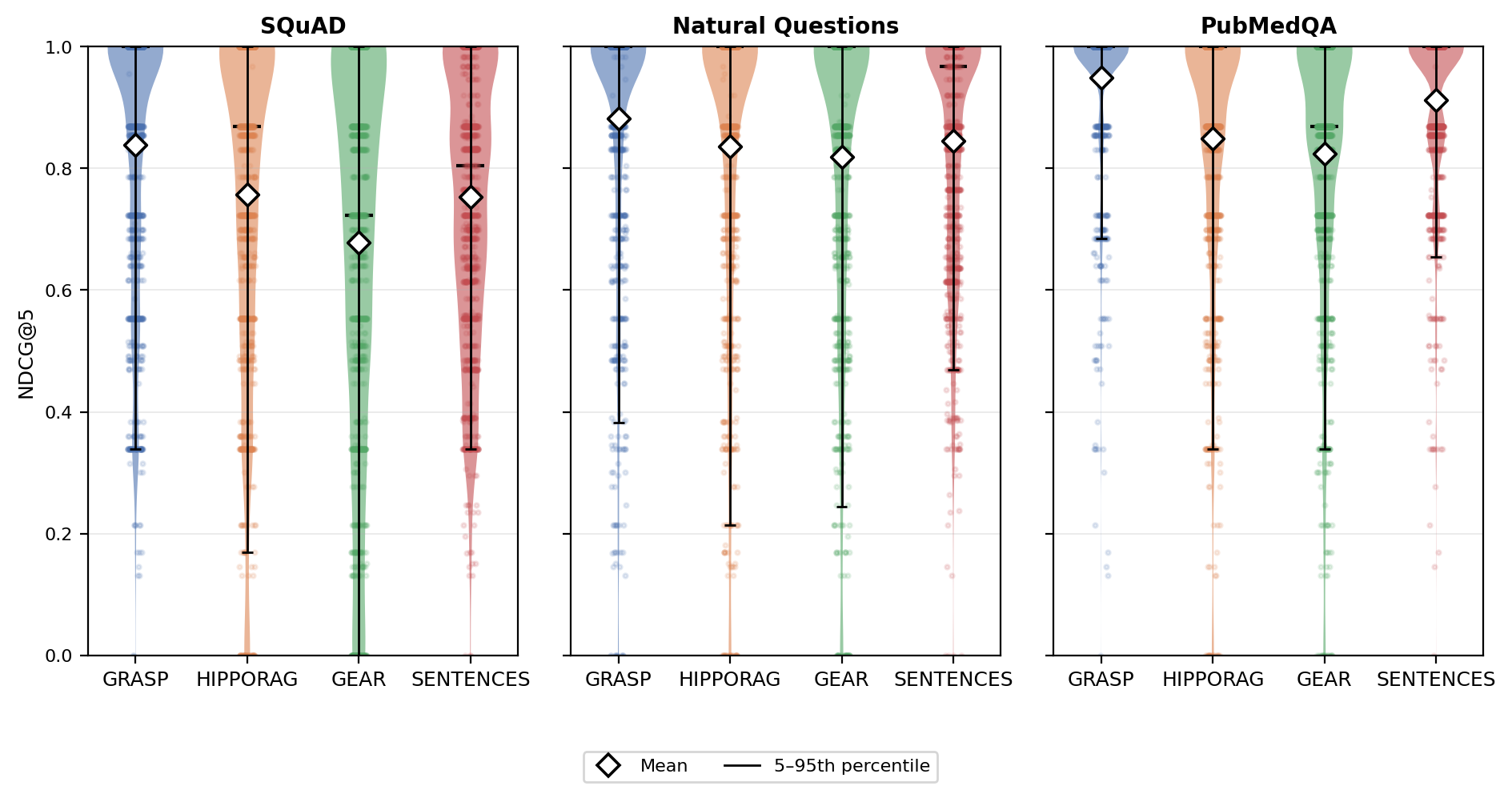}
    \caption{
        \textbf{NDCG@5 for retrieval unit discriminability}, measured by
        ranking all 1,000 questions' extracted units in a shared pool per
        method and treating gold units as relevant. Propositions
        achieve the highest discriminability across all three benchmarks.
        The sentence baseline — natural-language units without atomic
        decomposition — substantially outperforms GeAR across all datasets
        and HippoRAG on PubMedQA, implicating triple compression as the
        primary bottleneck. Propositions improve further upon sentences,
        showing that atomic decomposition provides an additional benefit.
    }
    \label{fig:embedding-similarity}
\end{figure*}

As shown in Figure~\ref{fig:embedding-similarity}, propositions achieve the highest discriminability across all three datasets, outperforming the strongest triple-based baseline (HippoRAG) by 8.1, 4.6, and 10.0 points on SQuAD, Natural Questions, and PubMedQA respectively. The sentence baseline illuminates the structure of this advantage. Sentences substantially outperform GeAR across all datasets (+7.6, +2.7, +8.9 points), and outperform HippoRAG on PubMedQA (+6.4 points) where dense biomedical phrasing resists clean triple decomposition — implicating triple compression itself, rather than extraction errors, as a primary driver of the gap. Propositions improve further upon sentences by 3.6--8.4 points across all three datasets, indicating that atomic decomposition into independently retrievable units provides discriminability benefits beyond natural-language format alone. Together, these results suggest that the embedding alignment advantage of propositions has two complementary sources: preserving natural-language form and decomposing passages into atomic, query-aligned units.

\subsubsection{Data Integrity Issues in HotpotQA}
\label{sec:hotpot-integrity}

HotpotQA is increasingly viewed as a near-solved dataset, diminishing its utility as a rigorous benchmark for evaluating agentic multi-hop reasoning. As highlighted by the creators of MuSiQue \citep{trivedi2022musique} and prior studies \citep{min2019compositional, chen2019understanding}, HotpotQA contains pervasive reasoning shortcuts that allow models to bypass genuine step-by-step deduction. Furthermore, because the dataset is widely circulated and based on general Wikipedia knowledge, its answers are deeply ingrained in the parametric memory of modern large language models. This allows models to "cheat" by relying on internalized facts rather than synthesizing evidence across retrieved passages. This parametric leakage largely explains our observation that zero-shot Gemini achieves a surprisingly competitive score on HotpotQA even in the \textbf{No Context} setting (Table \ref{tab:non_agentic}), whereas its closed-book performance sharply degrades on more rigorously constrained datasets like MuSiQue.

\begin{table*}[t]
  \centering
  \begin{tabular}{l ccc ccc}
    \toprule
    & \multicolumn{3}{c}{\textbf{Retrieval Setting}} & \multicolumn{3}{c}{\textbf{Extended context Setting}} \\
    \cmidrule(lr){2-4} \cmidrule(lr){5-7}
    & MuSiQue & 2Wiki & HotpotQA & MuSiQue & 2Wiki & HotpotQA \\
    \midrule
    \# queries            & 1{,}000 & 1{,}000 & 1{,}000 & 200 & 200 & 200 \\
    \# passages           & 11{,}656 & 6{,}119 & 9{,}811 & 2{,}218 & 1{,}986  & 1{,}722 \\    Avg. \# tokens & 2{,}344 & 903 & 1{,}300     & 15{,}976 & 13{,}148 & 7{,}227\\
    \bottomrule
  \end{tabular}
    \caption{Dataset statistics for both evaluation settings. Retrieval splits from HippoRAG2 \citep{gutierrez2025from}; Extended context splits from LongBench \citep{bai2024longbench}. Token count is derived from \texttt{gemini-3-flash}.}
    \label{tab:dataset_stats}
\end{table*}

\subsubsection{GRASP Structural Ablations}
\label{sec:graph-structural-ablation}

\begin{figure*}[h!]
  \centering
    \includegraphics[width=\textwidth]{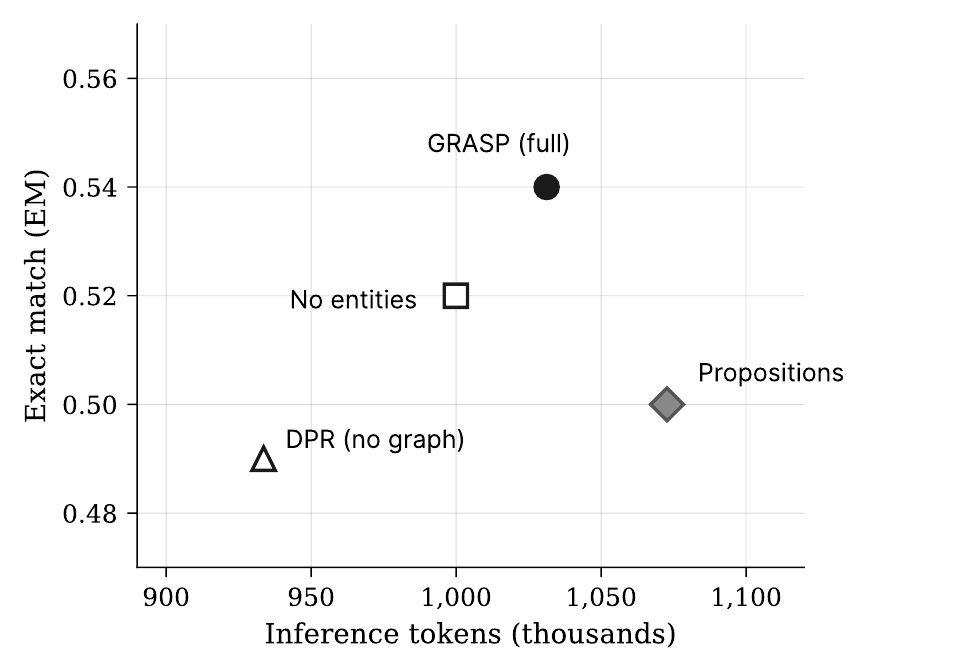}
  \caption{Ablation results of the varying retrieval structures used by GRASP's agent framework. Results reported on a 100-question random subset of MuSiQue.}
  \label{fig:grasp-ablations}
\end{figure*}

We ablate the retrieval structure underlying GRASP's sub-agent search by comparing the full pipeline as described in section~\ref{sec:method} against three variants. Dense Passage Retrieval (DPR) bypasses the propositional graph entirely, embedding and retrieving passages directly with a language encoder; this is the cheapest configuration but yields the lowest EM (0.49), indicating encoding passages alone is suboptimal. Removing entity selection retains the propositional layer but eliminates the entity-level filtering step, resulting in a 2-point EM drop relative to the full system at comparable token cost; this confirms that the entity layer provides an efficient narrowing of the search space. Finally, returning propositions directly via an adaptive-k mechanism \citep{taguchi2025}, rather than voting over propositions to retrieve full passages, produces the largest accuracy loss (EM 0.50, F1 0.65) despite consuming the most tokens, suggesting that grounding the agent's reasoning in complete passage context is important for multi-hop QA. This also indicates that propositions are useful retrieval units, but suboptimal for supplying the LLM context for answering questions fully. Taken together, these results validate that each layer in GRASP's index contribute meaningfully to downstream QA performance.

\subsubsection{Robustness of Success Economy to the Difficulty Judge}
\label{sec:success-economy-judge}

Success economy (Eq.~\ref{eq:success-economy}) requires only a per-question difficulty weight $w_i = -\log_2(r_i)$, so it is not tied to the backbone that produces the answers. To test whether our conclusions depend on the model used to estimate $r_i$, we recomputed the MuSiQue weights with \texttt{Claude Sonnet~5} in place of \texttt{gemini-3-flash-preview}, using the same protocol described in \S~\ref{sec:implementation}. Token usage $\mathcal{T}_i$ and correctness $\mathrm{EM}_i$ are unchanged. Every method still runs on the Gemini backbone, and only the weights $w_i$ differ.

\begin{table}[h]
  \centering
  \small
  \begin{tabular}{lcc}
    \toprule
    Method & $C_w$ (Gemini $w_i$) & $C_w$ (Sonnet $w_i$) \\
    \midrule
    \ours            & \textbf{12.6k} & \textbf{13.4k} \\
    GeAR             & 15.5k          & 16.5k          \\
    IRCoT+HippoRAG2  & 26.7k          & 27.7k          \\
    \bottomrule
  \end{tabular}
  \caption{Success economy on MuSiQue (retrieval setting) under difficulty weights estimated by Gemini and by Claude Sonnet~5. All methods use the Gemini backbone; lower is better.}
  \label{tab:success-economy-judge}
\end{table}

Table~\ref{tab:success-economy-judge} shows that the Sonnet weights raise $C_w$ by 0.8--1.0k for every method, since Sonnet answers more questions correctly and thus assigns lower weights. The ranking and the relative gaps between methods do not change. \ours\ remains the most token-efficient method, ahead of GeAR by about 19\% and ahead of IRCoT+HippoRAG2 by about 52\% under both judges. Per question, the Gemini and Sonnet correctness fractions $r_i$ agree moderately (Pearson $= 0.549$). The absolute weights therefore depend on the judge, but the comparisons success economy is used for in the paper do not.

\subsection{Storage Footprint and Graph Complexity}
\label{sec:storage-footprint}

We ran an indexing analysis of GRASP, HippoRAG, and GeAR across all benchmarks
in the retrieval setting, using identical hardware and capping each method at
$10$ indexing workers for a fair comparison. Table~\ref{tab:storage} reports
construction time, graph size, and storage footprint. We used \texttt{gemini-3-flash} as the extractor and \texttt{gemini-embedding-001} as the embedding model.

\begin{table*}[t]
  \centering
  \small
  \setlength{\tabcolsep}{4pt}
  \renewcommand{\arraystretch}{0.95}
  \begin{tabular}{ll r r r r r r}
    \toprule
    \textbf{Dataset} & \textbf{Method}
      & \textbf{Time (min)} & \textbf{Nodes} & \textbf{Entities}
      & \textbf{Edges} & \textbf{Props.} & \textbf{Store (GB)} \\
    \midrule
    \multirow{3}{*}{MuSiQue}
      & GRASP    & \textbf{38.8}  & 167{,}074 & 55{,}586 & 387{,}651   & 101{,}267 & \textbf{1.17} \\
      & HippoRAG & 119.6 & 88{,}319  & 76{,}663 & 4{,}488{,}183 & --        & 5.13 \\
      & GeAR     & 112.4 & 188{,}332 & 68{,}217 & 108{,}459   & --        & 1.67 \\
    \midrule
    \multirow{3}{*}{2Wiki}
      & GRASP    & \textbf{20.5} & 86{,}791  & 31{,}567 & 193{,}433   & 49{,}257 & \textbf{0.57} \\
      & HippoRAG & 31.4 & 46{,}190  & 40{,}071 & 2{,}286{,}454 & --       & 2.59 \\
      & GeAR     & 32.4 & 98{,}732  & 37{,}446 & 55{,}167    & --       & 0.85 \\
    \midrule
    \multirow{3}{*}{HotpotQA}
      & GRASP    & \textbf{34.8} & 156{,}979 & 53{,}533 & 366{,}896   & 94{,}165 & \textbf{1.09} \\
      & HippoRAG & 57.8 & 81{,}961  & 72{,}152 & 3{,}939{,}246 & --       & 4.76 \\
      & GeAR     & 56.7 & 176{,}664 & 65{,}272 & 101{,}581   & --       & 1.54 \\
    \bottomrule
  \end{tabular}
  \caption{Storage footprint and graph size of GRASP versus HippoRAG and GeAR in
  the retrieval setting, at $10$ indexing workers on identical hardware. Best per
  dataset is \textbf{bolded}.}
  \label{tab:storage}
\end{table*}

GRASP has the smallest storage footprint on all three datasets. We attribute this
to GRASP encoding only proposition nodes: these embeddings account for the
majority of its storage cost (1.1 of 1.17\,GB on MuSiQue), and GRASP stores no
embeddings for graph edges. This follows from its deterministic traversal from the
proposition layer to both the entity and passage layers, which requires no learned
edge representations. By contrast, HippoRAG and GeAR use \texttt{(subject,
predicate, object)} triples as retrieval units, embedding relational strings
rather than nodes alone and thereby storing far more vectors; HippoRAG's triple
graph reaches over four million edges on MuSiQue, driving its $5.13$\,GB footprint. Furthermore, HippoRAG embeds each \textit{passage} as well to support DPR as a fallback during retrieval.

GRASP also indexes faster than both baselines under these conditions, though we
acknowledge construction time is subject to engineering factors in a deployed
setting. Storage could be reduced further by lowering the embedding
dimensionality, which we set to $d = 3072$ for all methods.

\subsection{Prompts}
\label{sec:appendix-prompts}

\label{sec:prompts}

All prompts used in GRASP are reproduced below in full. Each prompt is labeled with the pipeline stage and corresponding section reference.

\begin{figure*}[t] %
    \centering
    \includegraphics[width=\textwidth]{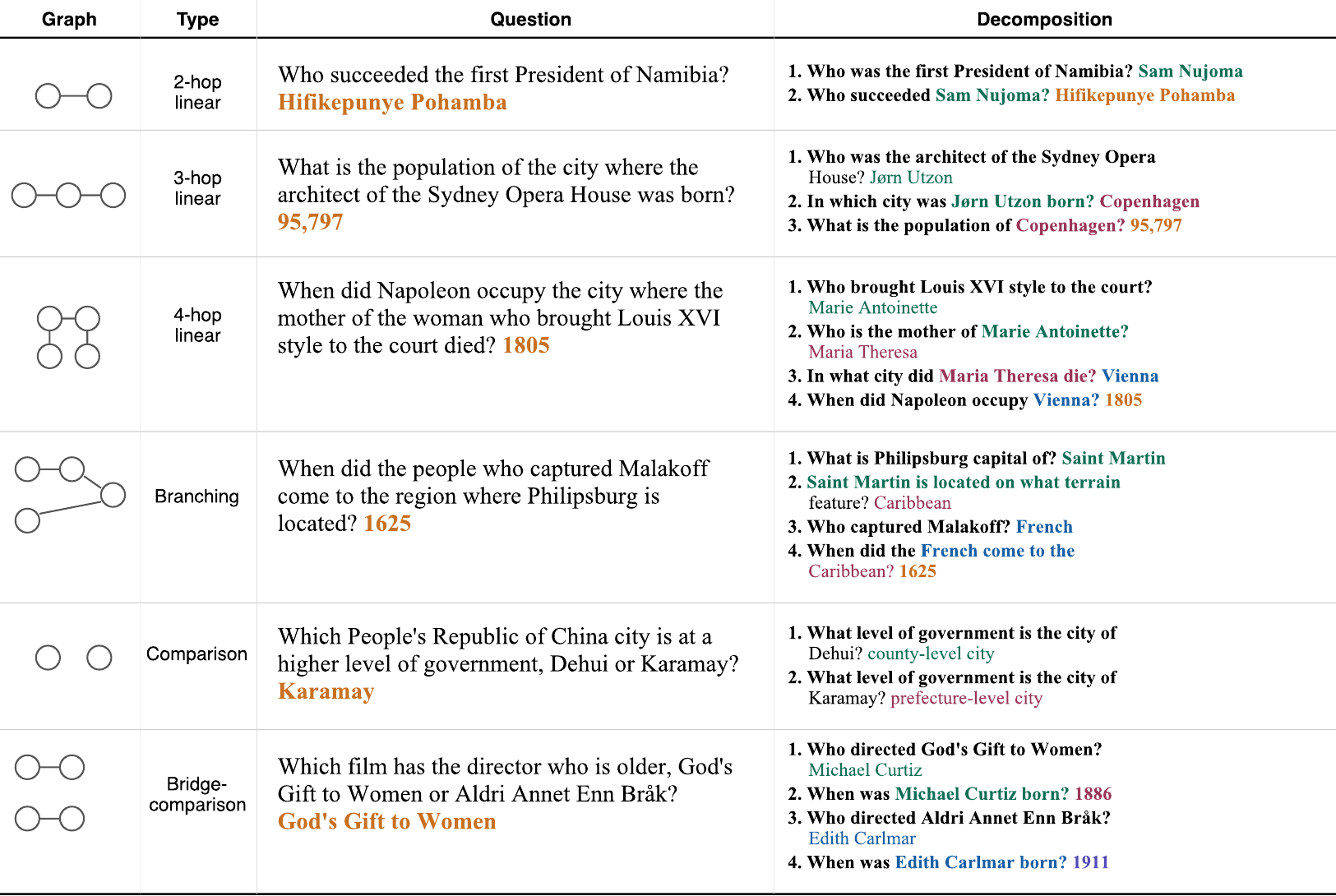}
    \caption{Breakdown of the multi-hop questions used as few-shot examples in our planner's prompt.}
    \label{fig:question-types}
\end{figure*}

\paragraph{Planner prompt (\S\ref{sec:agent-planning}).}
The planner decomposes multi-hop questions into dependency-ordered sub-questions using \texttt{\#N} placeholders. Few-shot examples cover linear chains, parallel branches, and comparison structures.

\begin{tcolorbox}[
    breakable,
    colback=gray!5,
    colframe=gray!60,
    boxrule=0.5pt,
    left=6pt, right=6pt, top=6pt, bottom=6pt
]
\ttfamily\fontsize{7}{9}\selectfont

You are a planner for a multi-hop question-answering system backed by a knowledge graph.

\vspace{4pt}
\textbf{Task}\\
Given a user question:\\
1. Write a \textbf{rational plan} --- a 1--3 sentence outline of the reasoning chain and key facts needed.\\
2. Produce an \textbf{ordered list of sub-questions}. Each will be answered by a research agent searching the knowledge graph.

\vspace{4pt}
\textbf{Rules}\\
- \textbf{No assumptions:} Never name or assume entities not explicitly stated in the question. Your sub-questions must \emph{find} them.\\
- \textbf{No redundant steps:} Every sub-question must retrieve new information from the knowledge graph. Do not create a step whose answer can be inferred from the question text or from a previous step's answer without a lookup.\\
- \textbf{Prefer direct search:} If an entity can be found directly (e.g., ``the university in state X''), search for it in one step rather than chaining through intermediate entities.\\
- \textbf{Dependencies:} Use \texttt{\#1}, \texttt{\#2}, \texttt{\#3}, etc.\ to reference answers from prior sub-questions. Never repeat context or use vague pronouns.\\
- \textbf{Parallel branches:} If the final answer requires two independent facts, retrieve them in separate sub-questions before combining in a later step.\\
- \textbf{Compound constraints:} If the question asks for an entity satisfying multiple constraints (e.g., ``who did X AND Y?''), resolve any unknown references first, then issue a single compound sub-question --- do not enumerate and intersect sets.\\
- \textbf{Maximum 4 sub-questions.} If your plan requires more, look for steps that can be combined or eliminated.

\vspace{4pt}
\textbf{Examples}

\vspace{2pt}
\textbf{Question:} ``Who succeeded the first President of Namibia?''\\
\textbf{Rational Plan:} I need to identify Namibia's first President and then find his/her successor.\\
\textbf{Sub-questions:}\\
1. Who was the first President of Namibia?\\
2. Who succeeded \#1?

\vspace{4pt}
\textbf{Question:} ``What is the population of the city where the architect of the Sydney Opera House was born?''\\
\textbf{Rational Plan:} I need to identify the architect of the Sydney Opera House, find their birth city, then retrieve that city's population.\\
\textbf{Sub-questions:}\\
1. Who was the architect of the Sydney Opera House?\\
2. In which city was \#1 born?\\
3. What is the population of \#2?

\vspace{4pt}
\textbf{Question:} ``When did Napoleon occupy the city where the mother of the woman who brought Louis XVI style to the court died?''\\
\textbf{Rational Plan:} I need to find who brought Louis XVI style to court, find that person's mother, find the city where the mother died, then find when Napoleon occupied that city. The chain is linear with four hops.\\
\textbf{Sub-questions:}\\
1. Who brought Louis XVI style to the court?\\
2. Who was the mother of \#1?\\
3. In what city did \#2 die?\\
4. When did Napoleon occupy \#3?

\vspace{4pt}
\textbf{Question:} ``When did the people who captured Malakoff come to the region where Philipsburg is located?''\\
\textbf{Rational Plan:} This requires two parallel branches --- finding the region where Philipsburg is located (via what it is capital of, then what terrain feature that is on) and who captured Malakoff --- before combining them in a final question.\\
\textbf{Sub-questions:}\\
1. What is Philipsburg capital of?\\
2. \#1 is located on what terrain feature?\\
3. Who captured Malakoff?\\
4. When did \#3 come to \#2?

\vspace{4pt}
\textbf{Question:} ``Which People's Republic of China city is at a higher level of government, Dehui or Karamay?''\\
\textbf{Rational Plan:} I need to find the government level of each city independently and then compare them. Both lookups are parallel.\\
\textbf{Sub-questions:}\\
1. What level of government is the city of Dehui?\\
2. What level of government is the city of Karamay?

\vspace{4pt}
\textbf{Question:} ``Where did Kent Patterson play his home games until he became a free agent?''\\
\textbf{Rational Plan:} I need to find which team Kent Patterson played for before becoming a free agent, then find where that team plays its home games.\\
\textbf{Sub-questions:}\\
1. Which team did Kent Patterson play for before becoming a free agent?\\
2. Where does \#1 play its home games?

\vspace{4pt}
\textbf{Question:} ``Which film has the director who is older, God's Gift to Women or Aldri Annet Enn Br\aa k?''\\
\textbf{Rational Plan:} I need to find the director of each film, then their birth dates, and compare. The two director lookups are independent but each requires a bridge through the film to the director's birth date.\\
\textbf{Sub-questions:}\\
1. Who directed God's Gift to Women?\\
2. When was \#1 born?\\
3. Who directed Aldri Annet Enn Br\aa k?\\
4. When was \#3 born?

\end{tcolorbox}

\paragraph{Query rewriting prompt (\S\ref{sec:subagent}).}
The sub-question is reformulated into a declarative search statement optimized for cosine similarity against propositions, along with keywords for BM25 sparse retrieval. This produces the inputs to step~\ding{182} in Figure~\ref{fig:architecture}.

\begin{tcolorbox}[
    breakable,
    colback=gray!5,
    colframe=gray!60,
    boxrule=0.5pt,
    left=6pt, right=6pt, top=6pt, bottom=6pt
]
\ttfamily\fontsize{7}{9}\selectfont

You are a dense retrieval query optimizer. Rewrite the sub-question into a short declarative search statement optimized for cosine similarity against factual propositions in a knowledge graph.

\vspace{4pt}
\textbf{Global Context}\\
Original Question: \{original\_question\}\\
Rational Plan: \{rational\_plan\}

\vspace{4pt}
\textbf{Current Step}\\
Context History: \{context\_history\}\\
Raw Sub-question: \{current\_sub\_question\}

\vspace{4pt}
\textbf{Rules}\\
1. \textbf{Exact entity substitution:} When replacing \#N, use the exact entity name from the Step N Answer. Do not paraphrase, generalize, or substitute a related entity. Strip citation tags (e.g., [ID: 123]) but preserve the entity name verbatim.\\
2. \textbf{No external knowledge:} Only use entity names that appear in the Raw Sub-question, Context History, or Original Question. Never inject names, spellings, or facts from your own knowledge.\\
3. \textbf{Self-contained statement:} Do not carry over relationships from prior steps (e.g., not ``John's spouse'' --- just the spouse's resolved name). The statement must stand alone.\\
4. \textbf{Declarative form:} Write a short declarative statement or noun phrase, as in an encyclopedia. No question words, no question marks. Lead with the resolved entity name when possible.\\
5. \textbf{Topical terms:} Include key terms from the Rational Plan that describe the information being sought (e.g., ``population'', ``founded'', ``spouse'') so the statement shares vocabulary with target propositions.

\vspace{4pt}
\textbf{Example}

\vspace{2pt}
Context History:\\
Step 1 Answer: The director of Oppenheimer is Christopher Nolan [ID: 293].

\vspace{2pt}
Raw Sub-question: What was the budget for the first movie directed by \#1?

\vspace{2pt}
Search Statement: The production budget of Christopher Nolan's first directed film.\\
Keywords: [``Christopher Nolan'']

\end{tcolorbox}

\paragraph{Entity selection prompt (\S\ref{sec:subagent}).}
Given candidate entities from hybrid search (step~\ding{182}), the agent selects which entities to traverse (step~\ding{183} in Figure~\ref{fig:architecture}). The reasoning context includes the observation state $O$ and previously visited nodes.

\begin{tcolorbox}[
    breakable,
    colback=gray!5,
    colframe=gray!60,
    boxrule=0.5pt,
    left=6pt, right=6pt, top=6pt, bottom=6pt
]
\ttfamily\fontsize{7}{9}\selectfont

Select the most relevant entity nodes to explore from a knowledge graph.

\vspace{4pt}
Reasoning Context: \{state\_context\}

\vspace{4pt}
Sub-question: \{sub\_question\}\\
Search statement: \{search\_statement\}

\vspace{4pt}
Candidate entities:\\
\{candidates\}

\vspace{4pt}
Choose one or more entity nodes whose propositions are most likely to contain information that answers the search statement. Select multiple nodes when they are all plausibly relevant (e.g., multiple entities mentioned in the search statement, or when the answer may require combining facts across nodes). Select only one when a single node clearly dominates.

\vspace{4pt}
Avoid selecting nodes listed under ``Already visited node IDs'' --- their propositions have already been explored.

\vspace{4pt}
Output the node\_ids (list of integers) of your selections and brief reasoning.

\end{tcolorbox}

\paragraph{Evidence evaluation prompt (\S\ref{sec:subagent}).}
After visiting entity nodes and observing their propositions, the agent evaluates whether accumulated evidence is sufficient (\textsc{done}) or whether a new search with a rewritten statement is needed (\textsc{query\_again}). This corresponds to step~\ding{185} in Figure~\ref{fig:architecture}. The grounding rules prevent the LLM from injecting parametric knowledge.

\begin{tcolorbox}[
    breakable,
    colback=gray!5,
    colframe=gray!60,
    boxrule=0.5pt,
    left=6pt, right=6pt, top=6pt, bottom=6pt
]
\ttfamily\fontsize{7}{9}\selectfont

You are evaluating evidence gathered from a knowledge graph to determine how to proceed with answering a question.

\vspace{4pt}
\textbf{== CURRENT STATE ==}\\
\{state\_block\}

\vspace{4pt}
\textbf{== NEW EVIDENCE ==}\\
\{new\_evidence\}

\vspace{4pt}
\textbf{Your task}

\vspace{2pt}
Analyze the new evidence in context of your current state. The visited nodes section summarizes what was learned in prior iterations --- use it alongside the new evidence above.

\vspace{4pt}
\textbf{Critical grounding rules:}

\vspace{2pt}
1. \textbf{Answer from the evidence, not from memory.} Your answer must be derived from the retrieved propositions and visited-node findings. Do not inject facts that are absent from the evidence. 

\vspace{2pt}
2. \textbf{The best answer in the evidence IS the answer.} If the question asks about a category (e.g., ``university,'' ``country,'' ``author'') and the evidence contains exactly one entity that fits that category --- even if it is not a perfect categorical match --- that entity is the answer. For example: if the question asks for a ``university'' and the only educational institution in the evidence is a Gymnasium, the answer is the Gymnasium. Do not reject the evidence in favor of searching for a ``better'' answer from your prior knowledge.

\vspace{2pt}
3. \textbf{Use reasoning to connect evidence, not to override it.} You may infer across propositions (e.g., chaining ``A founded B'' and ``B is located in C'' to conclude ``A founded something in C''). You may NOT substitute your own factual knowledge for missing evidence.

\vspace{4pt}
Can the sub-question be answered --- directly or by inference --- from what has been gathered? Choose exactly one of two actions:

\vspace{4pt}
\textbf{DONE} --- the answer is present or can be inferred

\vspace{2pt}
Use this when the evidence directly states or strongly implies the answer. \textbf{If the answer can be reasonably concluded or inferred from the available evidence, even without an explicit statement, treat this as DONE.}

\vspace{2pt}
- action: ``DONE''\\
- answer: State the answer with evidence ID citations, e.g.\ ``Christopher Nolan directed Inception [ID: 42].''\\
~~If multiple entities each validly answer the sub-question, use a numbered format: ``CANDIDATES: 1) Entity Name [ID: 42] 2) Other Entity [ID: 57]''. Do not add explanatory text after the candidate list.\\
- supporting\_prop\_ids: Evidence IDs involved in your reasoning.\\
- node\_findings: A 1--2 sentence summary of what was learned from the newly visited nodes, relative to the subquestion.\\
- missing\_information: ``resolved''

\vspace{4pt}
\textbf{QUERY\_AGAIN} --- the answer is definitively not recoverable from current evidence

\vspace{2pt}
Use this \textbf{only} when there is a clear gap that cannot be bridged by reasoning over the current evidence. Before choosing this, ask: ``Is the answer truly unrecoverable, or am I missing an inferential path?''

\vspace{2pt}
\textbf{If returning QUERY\_AGAIN, provide a meaningfully different search statement --- broaden to related entities, shared attributes, or co-occurrence rather than repeating the same angle.}

\vspace{2pt}
Note: source passages will be retrieved automatically after QUERY\_AGAIN --- you do not need to request them.

\vspace{2pt}
- action: ``QUERY\_AGAIN''\\
- answer: Best partial answer or intermediate conclusion so far. Describe what you can establish and what specific fact is missing. Use ``'' only if the evidence is entirely irrelevant.\\
- new\_search\_statement: A declarative statement (NOT a question) targeting the missing information.\\
- keywords: Proper nouns and named entities from the search statement for fulltext search. Group compound names (e.g.\ ``Christopher Nolan''). Exclude verbs, question words, and generic nouns.\\
- node\_findings: A 1--2 sentence summary of what was learned from the newly visited nodes.\\
- missing\_information: What specific information you still need to find.

\vspace{2pt}
Output your chosen action, reasoning, and all relevant fields.

\end{tcolorbox}

\paragraph{Answer synthesis prompt (\S\ref{sec:subagent}).}
After all sub-agents complete, the chain of question--answer pairs is passed to this module, which traces the reasoning chain step-by-step and produces a concise final answer. The ``Chain of Truth'' guideline enforces explicit logical connections between research steps.

\begin{tcolorbox}[
    breakable,
    colback=gray!5,
    colframe=gray!60,
    boxrule=0.5pt,
    left=6pt, right=6pt, top=6pt, bottom=6pt
]
\ttfamily\fontsize{7}{9}\selectfont

You are a synthesis assistant. Your objective is to provide a definitive answer to the user's question based on the provided research steps.

\vspace{4pt}
\textbf{Guidelines}

\vspace{2pt}
1. \textbf{Research-First Reasoning:} Ground your answer in the provided research steps. You may use reasoning to bridge small inferential gaps across steps, but prefer cited evidence over unsupported claims.

\vspace{2pt}
2. \textbf{Citation-First Rationale:} Every sentence in your rationale must cite the specific Research Step it draws from (e.g., ``According to Step \#1...'').

\vspace{2pt}
3. \textbf{The ``Chain of Truth'':} Trace the logic from the first step to the last. Do not skip steps. If Step 1 identifies a person, and Step 2 identifies their birth year, show that connection explicitly.

\vspace{2pt}
4. \textbf{Extract Concrete Values:} Prefer exact values (dates, names, numbers) over descriptive phrases. If the evidence contains ``following the 1981 export restraints,'' answer ``1981'' rather than paraphrasing with a relative expression.

\vspace{2pt}
5. \textbf{Primary Name Only:} When a research step identifies an answer with a parenthetical alias or ``also known as'' qualifier, use the primary name --- the name stated first, before any parenthetical --- not the alias.

\vspace{2pt}
6. \textbf{Concise Answers:} Your final answer should be as specific and brief as possible --- ideally no more than 6 words.

\vspace{4pt}
\textbf{Output Format}\\
You must follow this structure exactly. You must end with the phrase ``So the answer is:''

\vspace{2pt}
Rationale:\\
- [Step \#1]: $\langle$Evidence from step 1$\rangle$\\
- [Step \#2]: $\langle$How evidence from step 2 connects to step 1$\rangle$\\
...\\
- [Conclusion]: $\langle$Logical summary of the findings$\rangle$\\
- So the answer is: $\langle$Final Concise Answer$\rangle$

\vspace{4pt}
\textbf{Input Format}

\vspace{2pt}
Research: \{research\} \

\vspace{2pt}
Q: \{original\_question\}\\
A:

\end{tcolorbox}

\paragraph{Joint extraction prompt (\S\ref{sec:graph_construction}).} The prompt processes multiple passages per call, extracting atomic propositions and typed entities simultaneously. Entity-to-proposition index references form the entity--proposition edges of the hierarchical graph.

\label{sec:joint-extraction-prompt}

\begin{tcolorbox}[
    breakable,
    colback=gray!5,
    colframe=gray!60,
    boxrule=0.5pt,
    left=6pt, right=6pt, top=6pt, bottom=6pt
]
\ttfamily\fontsize{7}{9}\selectfont

You are an NLP assistant that jointly extracts propositions and typed entities from multiple passages in a single pass.

You are given a numbered list of passages. Process each passage \textbf{independently}, returning results in the same order.

\vspace{4pt}
For each passage:\\
1. \textbf{Decompose} Content into atomic, self-contained propositions. Replace all pronouns with canonical names using the Document Title. Split compound sentences; condense multi-attribute statements.\\
2. \textbf{Extract} unique named entities (people, organizations, locations, proper nouns, specific years) from the propositions. Use Title Case canonical names (e.g ``United States'' not ``The US''). Assign specific types (e.g., ``Sculptor'' not ``Person'').

\vspace{4pt}
\textbf{Output Format}

\vspace{2pt}
Passage [N]:\\
Propositions:\\
{[0]} [proposition text]\\
{[1]} [proposition text]\\
...

\vspace{2pt}
Entities:\\
{[Name]}|[Type]|[proposition indices]\\
...

\vspace{2pt}
Repeat for each passage. Proposition indices are 0-based and reset per passage.

\textbf{Example}

\vspace{6pt}
\textbf{Input:}\\
Passage [0]:\\
Document Title: Easter Hare\\
Content: The earliest evidence for the Easter Hare (Osterhase) was recorded in south-west Germany in 1678 by the professor of medicine Georg Franck von Franckenau, but it remained unknown in other parts of Germany until the 18th century. Scholar Richard Sermon writes that ``hares were frequently seen in gardens in spring.''

\vspace{6pt}
\textbf{Output:}\\
Passage [0]:\\
Propositions:\\
{[0]} The earliest evidence for the Easter Hare was recorded in south-west Germany in 1678 by Georg Franck von Franckenau.\\
{[1]} Georg Franck von Franckenau was a professor of medicine.\\
{[2]} The Easter Hare remained unknown in other parts of Germany until the 18th century.\\
{[3]} Richard Sermon was a scholar.\\
{[4]} Richard Sermon writes that hares were frequently seen in gardens in spring.\\
{[5]} The Easter Hare is also known as Osterhase.

\vspace{4pt}
Entities:\\
Easter Hare|Folklore Figure|0 2 5\\
Germany|Country|0 2\\
1678|Year|0\\
Georg Franck von Franckenau|Professor of Medicine|0 1\\
Richard Sermon|Scholar|3 4\\
Osterhase|Folklore Figure|5

\end{tcolorbox}

\definecolor{graspblue}{RGB}{76,114,176}
\definecolor{agentgold}{RGB}{180,140,40}
\definecolor{toolgray}{RGB}{230,232,235}
\definecolor{evidencebg}{RGB}{245,247,250}
\definecolor{donegold}{RGB}{180,140,40}
\definecolor{queryorange}{RGB}{200,100,40}
\definecolor{answerbg}{RGB}{235,245,235}

\tcbset{
  graspbox/.style={
    enhanced, breakable,
    colframe=graspblue, colback=white,
    fonttitle=\bfseries\small,
    coltitle=white, colbacktitle=graspblue,
    attach boxed title to top left={yshift=-2mm, xshift=4mm},
    boxed title style={sharp corners, colback=graspblue},
    sharp corners, boxrule=0.8pt,
    left=6pt, right=6pt, top=8pt, bottom=6pt,
  },
  agentbox/.style={
    enhanced, breakable,
    colframe=agentgold!80!black, colback=white,
    fonttitle=\bfseries\small,
    coltitle=white, colbacktitle=agentgold!80!black,
    attach boxed title to top left={yshift=-2mm, xshift=4mm},
    boxed title style={sharp corners},
    sharp corners, boxrule=0.8pt,
    left=6pt, right=6pt, top=8pt, bottom=6pt,
  },
  evidencebox/.style={
    enhanced, colback=evidencebg, colframe=gray!40,
    sharp corners, boxrule=0.5pt,
    left=5pt, right=5pt, top=4pt, bottom=4pt,
    fontupper=\small\itshape,
  },
  actiondone/.style={
    enhanced, colback=answerbg, colframe=green!50!black,
    sharp corners, boxrule=0.5pt,
    left=5pt, right=5pt, top=4pt, bottom=4pt,
    fontupper=\small,
  },
  actionquery/.style={
    enhanced, colback=orange!8, colframe=queryorange,
    sharp corners, boxrule=0.5pt,
    left=5pt, right=5pt, top=4pt, bottom=4pt,
    fontupper=\small,
  },
}

\newcommand{\toolname}[1]{\texttt{\small #1}}
\newcommand{\fieldlabel}[1]{\textit{\small\color{gray!70!black}#1:}\ }
\newcommand{\proptext}[1]{\begin{tcolorbox}[evidencebox]``#1''\end{tcolorbox}}
\newcommand{\stepnumber}[1]{\textbf{Step #1}}

\subsection{GRASP End-to-End Worked Example (3-Hop)}

We trace a complete GRASP run on a 3-hop MuSiQue question to illustrate
planning, iterative subagent reasoning, and evidence-grounded synthesis.

\begin{tcolorbox}[graspbox, title={Question}]
  When was the Palau de la Generalitat constructed in the city where
  Martin from the region where Perdiguera is located died?

  \medskip
  \fieldlabel{Ground-truth answer} \textbf{built in the 15th century}
\end{tcolorbox}

\bigskip

\begin{tcolorbox}[graspbox, title={Planner}]

  \fieldlabel{Rational plan}
  Identify the region where Perdiguera is located, find the person named
  Martin from that region, determine where he died, and then find the
  construction date of the Palau de la Generalitat in that city.

  \medskip
  \fieldlabel{Decomposed sub-questions}
  \begin{enumerate}[leftmargin=1.4em, itemsep=2pt, topsep=2pt]
    \item In which region is Perdiguera located?
    \item Who is the person named Martin from \emph{\#1} and in which
          city did he die?
    \item When was the Palau de la Generalitat in \emph{\#2} constructed?
  \end{enumerate}
\end{tcolorbox}

\bigskip

\begin{tcolorbox}[agentbox, title={Subagent 1}]

  \fieldlabel{Given sub-question} In which region is Perdiguera located?

  \fieldlabel{Rewritten query}
  The administrative region or autonomous community where the municipality
  of Perdiguera is located.

  \medskip
  \noindent\rule{\linewidth}{0.3pt}

  \smallskip
  \toolname{traversal} \textbf{(iteration 1)}

  \smallskip
  \fieldlabel{Search statement}
  The administrative region or autonomous community where the municipality
  of Perdiguera is located.

  \fieldlabel{Keywords} Perdiguera, region, location

  \smallskip
  \fieldlabel{Top candidate nodes} Perdiguera, Balearic Islands, Asturias, Spain, Basque Country

  \smallskip
  \fieldlabel{Retrieved passage}
  \proptext{Perdiguera is a municipality located in the province of
    Zaragoza, \textbf{Aragon}, Spain.  According to the 2009 census
    (INE), the municipality has a population of 662 inhabitants.}

  \noindent\rule{\linewidth}{0.3pt}

  \smallskip
  \toolname{evaluation}

  \begin{tcolorbox}[actiondone]
    \textbf{Action:} \textsc{Done}\\
    \textbf{Reasoning:} The passage explicitly states that Perdiguera
    is located in Aragon.  In Spanish geography, Aragon is the autonomous
    community.\\
    \textbf{Answer:} Perdiguera is located in the autonomous community of
    \textbf{Aragon}.
  \end{tcolorbox}

  \medskip
  \fieldlabel{Subagent output}
  \textbf{Aragon} \hfill \textit{(5.1 s)}
\end{tcolorbox}

\bigskip

\begin{tcolorbox}[agentbox,
  title={Subagent 2}]

  \fieldlabel{Given sub-question} Who is the person named Martin from \emph{\#1} and in which city did he die?

  \fieldlabel{Rewritten query}
  The person named Martin from Aragon and the city where he died.

  \medskip
  \noindent\rule{\linewidth}{0.3pt}

  \smallskip
  \toolname{traversal} \textbf{(iteration 1)}

  \fieldlabel{Search statement}
  The person named Martin from Aragon and the city where he died.

  \fieldlabel{Keywords} Martin, Aragon, city of death, died

  \fieldlabel{Top candidate nodes} Aragon, Martin, Martin of Aragon,
  Crown of Aragon, 1162

  \fieldlabel{Selected node} Martin of Aragon

  \fieldlabel{Retrieved passage}
  \proptext{Martin the Humane (29 July 1356 – 31 May 1410), also called
    the Elder and the Ecclesiastic, was King of Aragon, Valencia,
    Sardinia and Corsica and Count of Barcelona from 1396 and King of
    Sicily from 1409.  He failed to secure the accession of his
    illegitimate grandson, Frederic, Count of Luna, and with him the
    rule of the House of Barcelona came to an end.}

  \smallskip
  \toolname{evaluation}

  \begin{tcolorbox}[actionquery]
    \textbf{Action:} \textsc{Query Again}\\
    \textbf{Reasoning:} Martin the Humane is identified as the relevant
    figure, but the current evidence gives only birth and death
    \emph{dates} — the city of death is not yet stated.\\
    \textbf{New query:} The death place and date of Martin the Humane of
    Aragon.\\
    \textbf{New keywords:} Martin the Humane, Aragon, death place
  \end{tcolorbox}

  \medskip
  \noindent\rule{\linewidth}{0.3pt}

  \smallskip
  \toolname{traversal} \textbf{(iteration 2)}

  \fieldlabel{Search statement}
  The death place and date of Martin the Humane of Aragon.

  \fieldlabel{Keywords} Martin the Humane, Aragon, death place

  \fieldlabel{Top candidate nodes} Aragon, Crown of Aragon, 1162, 1347, 1409

  \fieldlabel{Selected node} Crown of Aragon

  \fieldlabel{Retrieved passages}
  \proptext{Crown of Aragon \textnormal{[\ldots]} Composite monarchy,
    confederation of kingdoms \textnormal{[\ldots]} Capital:
    \textbf{Barcelona} \textnormal{[\ldots]} 1162–1716.}

  \proptext{The Treaty of Barcelona was signed on 19 January 1493
    between France and the Crown of Aragon.}

  \smallskip
  \toolname{evaluation}

  \begin{tcolorbox}[actiondone]
    \textbf{Action:} \textsc{Done}\\
    \textbf{Reasoning:} Martin the Humane is identified as King of Aragon
    (d.\ 31 May 1410).  The Crown of Aragon evidence associates
    \textbf{Barcelona} as the capital and primary seat of the monarchy.
    The Treaty of Barcelona further corroborates this.\\
    \textbf{Answer:} Martin the Humane (Martin of Aragon) died in
    \textbf{Barcelona}.
  \end{tcolorbox}

  \medskip
  \fieldlabel{Subagent output}
  \textbf{Barcelona} \hfill \textit{(9.9 s, 2 traversal iterations)}
\end{tcolorbox}

\bigskip

\begin{tcolorbox}[agentbox,
  title={Subagent 3}]

  \fieldlabel{Given sub-question} When was the Palau de la Generalitat in \emph{\#2} constructed?

  \fieldlabel{Rewritten query}
  The construction dates and historical building period of the Palau de
  la Generalitat in Barcelona.

  \medskip
  \noindent\rule{\linewidth}{0.3pt}

  \smallskip
  \toolname{traversal} \textbf{(iteration 1)}

  \fieldlabel{Search statement}
  The construction dates and historical building period of the Palau de
  la Generalitat in Barcelona.

  \fieldlabel{Keywords} Palau de la Generalitat, Barcelona, constructed,
  construction

  \fieldlabel{Top candidate nodes} Palau de la Generalitat, Barcelona,
  Josep Dom\`enech i Estap\`a, Palau G\"uell, Palau Montaner

  \fieldlabel{Retrieved passage}
  \proptext{[\ldots] Partial survivals of former royal residences include
    the Doge's Palace of Venice, the \textbf{Palau de la Generalitat in
    Barcelona, built in the 15th century} for the kings of Aragon, or
    the famous Conciergerie, former palace of the kings of France, in
    Paris.}

  \noindent\rule{\linewidth}{0.3pt}

  \smallskip
  \toolname{evaluation}

  \begin{tcolorbox}[actiondone]
    \textbf{Action:} \textsc{Done}\\
    \textbf{Reasoning:} The passage explicitly states the Palau de la
    Generalitat in Barcelona was built in the 15th century for the kings
    of Aragon.  This directly answers the sub-question.\\
    \textbf{Answer:} The Palau de la Generalitat in Barcelona was built
    in the \textbf{15th century}.
  \end{tcolorbox}

  \medskip
  \fieldlabel{Subagent output}
  \textbf{15th century} \hfill \textit{(5.0 s)}
\end{tcolorbox}

\bigskip

\begin{tcolorbox}[graspbox, title={Final Answer}]

  \fieldlabel{Rationale}
  \begin{enumerate}[leftmargin=1.4em, itemsep=3pt, topsep=3pt]
    \item Perdiguera is located in the autonomous community of
          \textbf{Aragon}. \textit{(Subagent 1)}
    \item The person named Martin from Aragon is \textbf{Martin the
          Humane}, who died in the city of \textbf{Barcelona}.
          \textit{(Subagent 2)}
    \item The Palau de la Generalitat, located in Barcelona, was
          constructed in the \textbf{15th century}. \textit{(Subagent 3)}
  \end{enumerate}

  \medskip
  \noindent\rule{\linewidth}{0.4pt}

  \medskip
  \fieldlabel{Prediction} \textbf{15th century} \quad
  {\color{green!60!black}\textbf{(Exact match)}}

  \smallskip
  \fieldlabel{Total duration} 22.8 s \quad
  \fieldlabel{LLM calls} 13

  \smallskip
  \fieldlabel{Input tokens} 9{,}620 \quad
  \fieldlabel{Output tokens} 1{,}373 \quad
  \fieldlabel{Total tokens} 10{,}993
\end{tcolorbox}

\end{document}